# Magnetic Immunoassays: A Review of Virus and Pathogen Detection Before and Amidst the Coronavirus Disease-19 (COVID-19)

Kai Wu[a,*], Renata Saha[a], Diqing Su[b], Venkatramana D. Krishna[c], Jinming Liu[a], Maxim C-J Cheeran[c,*] and Jian-Ping Wang[a,*]

[a] Department of Electrical and Computer Engineering, University of Minnesota, Minneapolis, MN 55455, United States

[b] Department of Chemical Engineering and Material Science, University of Minnesota, Minneapolis, MN 55455, United States

[c] Department of Veterinary Population Medicine, University of Minnesota, St. Paul, MN 55108, United States

* E-mails: wuxx0803@umn.edu (K. W.), cheeran@umn.edu (M. C-J. C), jpwang@umn.edu (J.-P. W.)

**Abstract**

The novel severe acute respiratory syndrome coronavirus 2 (SARS-CoV-2), which causes coronavirus disease 2019 (COVID-19), is a threat to the global healthcare system and economic security. As of July 2020, no specific drugs or vaccines are yet available for COVID-19, fast and accurate diagnosis for SARS-CoV-2 is essential in slowing down the spread of COVID-19 and for efficient implementation of control and containment strategies. Magnetic immunoassay is a novel and emerging topic representing the frontiers of current biosensing and magnetics areas. The past decade has seen rapid growth in applying magnetic tools for biological and biomedical applications. Recent advances in magnetic materials and nanotechnologies have transformed current diagnostic methods to nanoscale and pushed the detection limit to early stage disease diagnosis. Herein, this review covers the literatures of magnetic immunoassay platforms for virus and pathogen detections, before COVID-19. We reviewed the popular magnetic immunoassay platforms including magnetoresistance (MR) sensors, magnetic particle spectroscopy (MPS), and nuclear magnetic resonance (NMR). Magnetic Point-of-Care (POC) diagnostic kits are also reviewed aiming at developing plug-and-play diagnostics to manage the SARS-CoV-2 outbreak as well as preventing future epidemics. In addition, other platforms that use magnetic materials as auxiliary tools for enhanced pathogen and virus detections are also covered. The goal of this review is to inform the researchers of diagnostic and surveillance platforms for SARS-CoV-2 and their performances.

# 1. Introduction

In December 2019, a cluster of severe pneumonia cases of unknown cause was reported in Wuhan, Hubei province, China.[1] A novel strain of coronavirus belonging to the same family of viruses that cause severe acute respiratory syndrome (SARS) and Middle East respiratory syndrome (MERS) was subsequently isolated from bronchoalveolar lavage fluid (BALF).[2,3] The virus was initially named 2019 novel coronavirus (2019-nCoV) and later renamed as severe acute respiratory syndrome coronavirus 2 (SARS-CoV-2).[4,5] The outbreak that began in China has rapidly expanded worldwide and on January 30, 2020 the World Health Organization (WHO) declared novel corona virus infection a Public Health Emergency of International Concern and the illness was named coronavirus disease 2019 (COVID-19). COVID-19 was declared as a pandemic by WHO on March 11, 2020 due to its rapid spread in various countries around the world. SARS-CoV-2 is an enveloped, positive-strand RNA virus with large RNA genome of ~30kb with genome characteristics similar to known coronaviruses.[6,7] The coronavirus genomic RNA encodes replication and transcription complex from a single large open reading frame (ORF1ab) and structural proteins of the virus.[8] The major structural proteins of corona virus are spike (S), envelope (E), membrane (M), and nucleocapsid (N).

There is currently no medication to treat COVID-19. Since clinical manifestation of COVID-19 ranges from mild flu-like symptoms to life threatening pneumonia and acute respiratory illness, it is essential to have proper diagnosis during early stage of infection for efficient implementation of control measures to slow down the spread of COVID-19.[9–11] Currently, real time reverse transcription polymerase chain reaction (RT-PCR) is the most widely used laboratory test for diagnosis of COVID-19. RT-PCR detects SARS-CoV-2 RNA and target different genomic regions of viral RNA.[12–14] Although RT-PCR is sensitive technique, they require expensive laboratory equipment, trained technicians to perform the test, and can take up to 48 hours to generate results. In addition, studies have found up to 30% false negative rate for RT-PCR early in the course of infection.[15–18] Several laboratories around the world are working on improving RT-PCR methods and to develop alternative molecular diagnostic platforms. Isothermal nucleic acid amplification that allow rapid amplification of target sequences at a single constant temperature are employed in several tests including ID NOW COVID-19 test from Abbott diagnostics. ID NOW is a rapid, point-of-care test that allow direct detection of viral RNA from the clinical sample without the need for RNA extraction. However, recent studies have found false negative rates ranging from 12-48% mainly due to inappropriate condition of sample transportation and inappropriate sample.[19–21] Moreover, this can test only one sample per run. Serological methods like enzyme-linked immunosorbent assay (ELISA) and lateral flow immunochromatography test that detect antibodies can be used to monitor immunity to infection and disease progression.[22] Although negative SARS-CoV-2 antibody results does not rule out COVID-19, serological assays will help in assessing previous exposure to SARS-CoV-2 in a population and therefore have a potential use in understanding the epidemiology of COVID-19. Currently available serological assays can detect IgM, IgG, or IgA antibodies to spike (S) or nucleocapsid (N) protein.[23–25] However, potential cross reactivity of SARS-CoV-2 antibodies with antibodies generated against other coronaviruses is a challenge in developing accurate serological test for COVID-19.[26]

Among other biosensing technologies, magnetic biosensors have attracted special attention in the past 20 years. Both surface-based and volume-based magnetic biosensors have been developed for the detection of viruses, pathogens, cancer biomarkers, metallic ions, etc.[27–36] In magnetic biosensors, the magnetic tags (usually magnetic nanoparticles (MNPs)) are functionalized with antibodies or DNA/RNA probes that can specifically bind to target analytes. The concentration of the target analytes is thus converted to the magnetic signals that are generated by these magnetic tags. Compared to optical, plasmonic, and electrochemical biosensors, magnetic biosensors exhibit low background noise since most of the biological environment is non-magnetic. The sensor signal is also less influenced by the types of the sample matrix, enabling accurate and reliable detection processes.[37] The number of published papers on magnetic biosensors is summarized in Figure 1, which indicates an increasing scientific interest on this topic.

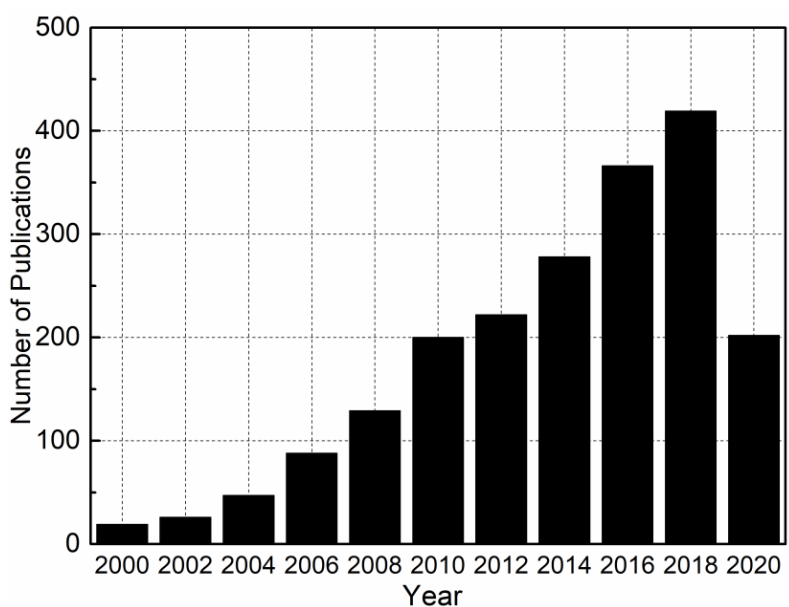

Figure 1. Number of publications on magnetic biosensors in the past 20 years as of July 8$^{th}$, 2020. The data is acquired from Web of Science core collection with the keywords "magnetic biosensors" and "magnetic biological sensors".

Most magnetic biosensors fall into several categories, namely magnetoresistance (MR) sensors, magnetic particle spectroscopy (MPS) platforms, and nuclear magnetic resonance (NMR) platforms. MR sensors are surface-based technologies which are sensitive to the stray field from the MNPs bound to the proximity of the sensor surface. The MR-based magnetic immunoassays are reviewed in Section 2, this kind of assay scheme is achieved by converting the binding events of MNPs (due to the presence of target analytes) to readable electric signals. On the contrary, MPS platforms (reviewed in Section 3) directly detect the dynamic magnetic responses of MNPs and thus, MNPs are the only signal sources and indicators for probing target analytes from non-magnetic mediums. NMR platforms (reviewed in Section 4) are using MNPs as contrast enhancers to introduce local magnetic field inhomogeneity and to disturb the precession frequency variations in millions of surrounding water protons. Thus, the high sensitivity NMR-based immunoassays intrinsically benefit from the MNP contrast agents. In addition, other immunoassay platforms that use magnetic materials as auxiliary tools to enhance the detection performances are also reviewed in Section 5. In this review, magnetic biosensors' application in virus and pathogen detection will be summarized and discussed based on the different working principle of the technologies.

## 2. Magnetoresistance (MR) Platforms
### 2.1 Magnetoresistance (MR)
Magnetoresistance (MR) was at first discovered by William Thompson who coined the term anisotropic magnetoresistance (AMR).[38] The physical observation of AMR shows that the resistivities of both Ni and Fe increase when the charge current is applied parallel to the magnetization and decrease when charge current is applied perpendicular to the magnetization.[39] This AMR effect originates from the spin orbit interactions and was experimentally and quantitatively demonstrated by Fert and Campbell.[40] However, the maximum resistance change recorded from AMR devices is only around 2 %, which renders it unsuitable for most applications. Regarding this, a detailed review of the AMR effect in thin films and bulk materials can be found in Ref. [39]. Herein, the AMR biosensors will not be discussed due to their limited applications in magnetic biosensing.

Giant magnetoresistance (GMR) was at first observed from the Fe/Cr multilayers grown with molecular beam epitaxy (MBE) by Albert Fert and Peter Grunberg.[41,42] These multilayers exhibit a resistance change significantly higher than the AMR devices. The GMR effect primarily exists in multilayer structures with alternating ferromagnetic and non-magnetic metallic layers. When the magnetizations of two adjacent ferromagnetic layers are parallel, the multilayers show low resistance and when magnetizations are anti-parallel, multilayers exhibit a high-resistance state. The industrial breakthrough for GMR discovery was made when Parkin *et al.* observed the GMR effect from DC sputtered multilayer structures.[43] Although the GMR effect was primarily observed in a thin film or layered system (see Figure 2(A)), it is also observed in other systems such as Co-Au, Co-Ag and Fe-Ag granular films.[44–48] GMR effect in granular films (see Figure 2(B)) is highly related to the spin dependent interfacial scattering, inter-particle coupling, and several are significant for biosensing purposes because of their capability to adapt to the shapes of different biomolecules.[49,50] In comparison to other types of sensors, the ability of flexible GMR sensors to respond to external magnetic field makes them a perfect candidate for wearable real-time body activity monitoring and evaluating drug delivery effectiveness. As no experimental demonstration on flexible MR-based detection of viruses/pathogens has been reported, further discussion on flexible GMR-based bio-detection is restricted in the subsequent sections.

Magnetic tunnel junctions (MTJs) have similar stack structure (see Figure 2(C)) to that of the GMR spin valves except that the adjacent ferromagnetic layers are separated by an insulating layer which is usually an oxide. In the earlier days, $AlO_x$ was used [51,52]. Later, this insulating layer was replaced by MgO material for smaller lattice mismatch and interface instability and thus, higher tunnel magnetoresistance (TMR) ratio [53,54]. The most important characteristic of a MTJ structure is its transfer curve as shown in Figure 2(D). In the transfer curve, two characteristics are of utmost importance: MR ratio and sensitivity. The physical characterization of the MR ratio is the rate of change in MR deice resistance along with varying magnetic field. Its sensitivity is measured by the slope of the transfer curve at an intensity of the magnetic field. In this regard, an interesting point to note is the tradeoff between the sensitivity and the linear magnetic field response range for MR sensors. A large linear response range in the transfer curve is attained with great ease in GMR sensors, although this comes with a compromise on the sensitivity. On the other hand, even though MTJ sensors possess high sensitivity, additional stack designs or supporting parts such as bias magnets are required to achieve high linearity.[55–57] Another factor which comes into play for all sensors in the nanoscale is the signal-to-noise ratio (SNR). Generally, MTJs show higher SNR than GMR sensors. However, the shot noise from the discontinuities in the conduction medium can cause the SNRs of MTJs to suffer.[58] With the advancing of thin film deposition and nanofabrication technologies, the TMR ratio has been increased dramatically during the past 20 years from ~20% to over 200%.[53,59–61]

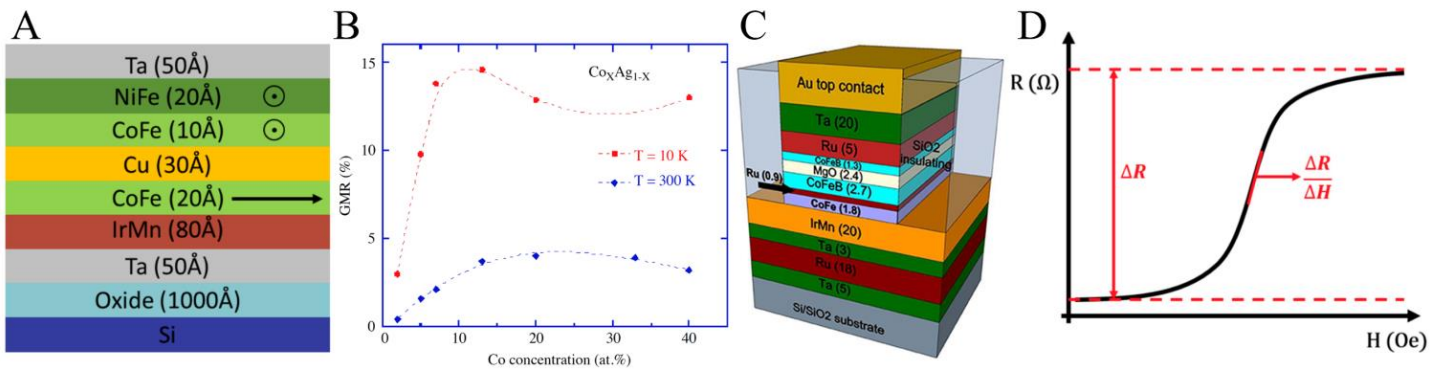

Figure 2. (A) A typical GMR stack structure used for biosensing. (B) Magnetoresistance of Co-Ag matrix, the evidence of granular GMR. (C) A typical MTJ structure used for biosensing. (D) A typical transfer curve of a MR sensor. (A) reprinted with permission from [59], Copyright (2019) IOP Publishing. (B) reprinted with permission from [47], Copyright (2006) Elsevier. (C) reprinted from [60] under the terms of the Creative Commons Attribution License (CC BY). (D) reprinted from [61] under the terms and conditions of the Creative Commons Attribution (CC BY) license.

## 2.2 Giant Magnetoresistance (GMR) Platform

Since Baselt *et al.* reported the first GMR-based biosensor using the Bead Array Counter (BARC) microarray, GMR-based biosensing has been attracting increasing attentions amongst the community.[62] This section reviews the GMR biosensors for detecting viruses and pathogens, and compares their limit of detections (LODs) and advantages over the existing biosensing tools. Take the sandwich immunoassay as an example (see Figure 3(A)), where the capture antibodies specifically targeting on analytes (such as antigens from viruses/pathogens) are pre-functionalized on the GMR sensor surface. Then biofluid samples are added and specific antibody-antigen bindings take place on the sensor surface. Usually a wash step is added to remove the unbound analytes from sensing areas. Then the detection antibody functionalized MNPs are added to the GMR sensing areas, forming the MNP – detection antibody – antigen – capture antibody complexes. Thus, the amount of MNPs captured to the proximity of sensor surface is directly proportional to the number of antigens in the testing sample. Furthermore, this sandwich immunoassay scheme significantly enhances the detection specificity. To attain the best performance, superparamagnetic MNPs are prevalently used to avoid clustering and sedimentation to sensor surfaces.

Krishna *et al.* reported a GMR benchtop system for the detection of H1N1 strain of the influenza A virus (IAV) within a concentration range of $10^3$ to $10^5$ TCID$_{50}$/mL.[63] Wu *et al.* reported a portable GMR biosensing device named Z-Lab (see Figure 3(B)) to detect IAV.[64] They achieved a LOD of 15 ng/mL for detecting H1N1 nucleoprotein (see Figure 3(C)) and a LOD of 125 TCID$_{50}$/mL for detecting purified H3N2 variant virus (H3N2v) from buffer solutions, with the overall assay time of less than 10 mins. Later, Su *et al.* reported the wash-free immunoassay scheme for detecting H1N1 and H3N2v from spiked nasal swab samples with a reported LOD of 250 TCID$_{50}$/mL.[27] This wash-free immunoassay approach allows for detections handled by non-technicians with minimum training requirements. Another group from Stanford University reported a similar GMR-based portable system for on-site bioassays (see Figure 3(D)). They reported the multiplexed assay of human immunoglobulin G and M (IgG and IgM) antibodies with sensitives down to 0.07 and 0.33 nanomolar, respectively. Figure 3(E) shows the real-time signals as measured by their portable device for detecting various concentrations of IgG over a 10 min measurement period.[65] Zhi *et al.* reported detection of Hepatitis B virus (HBV) using a GMR biochip integrated with microfluidic channel with a detection sensitivity of 200 IU/mL for HBV DNA molecules.[66] In their work, the integration of a microfluidic channel increased the ease of handling smaller sample volumes on the sensing area. A good follow-up of this work with significantly improved LOD down to 10 copies of target HBV DNA molecules has been reported.[67] GMR platforms have also been reported for bacteria detections. For instance, Sun *et al.* reported the detection of *Escherichia coli O157H:H7* antigen using the GMR biosensing scheme with a reported LOD of 100 colony forming unit (CFU)/mL.[68] Gupta *et al.* reported the detection of *Mycobacterium tuberculosis* specific antigen - ESAT-6 using the GMR scheme and reported a LOD of 1 pM.[69]

The key take-away point here is that several experimental demonstrations of the magnetic assays for virus detection based on GMRs and the reported LOD indicate that GMR-based bioassay is one of the promising candidates for onsite, rapid, and sensitive detection of COVID-19.

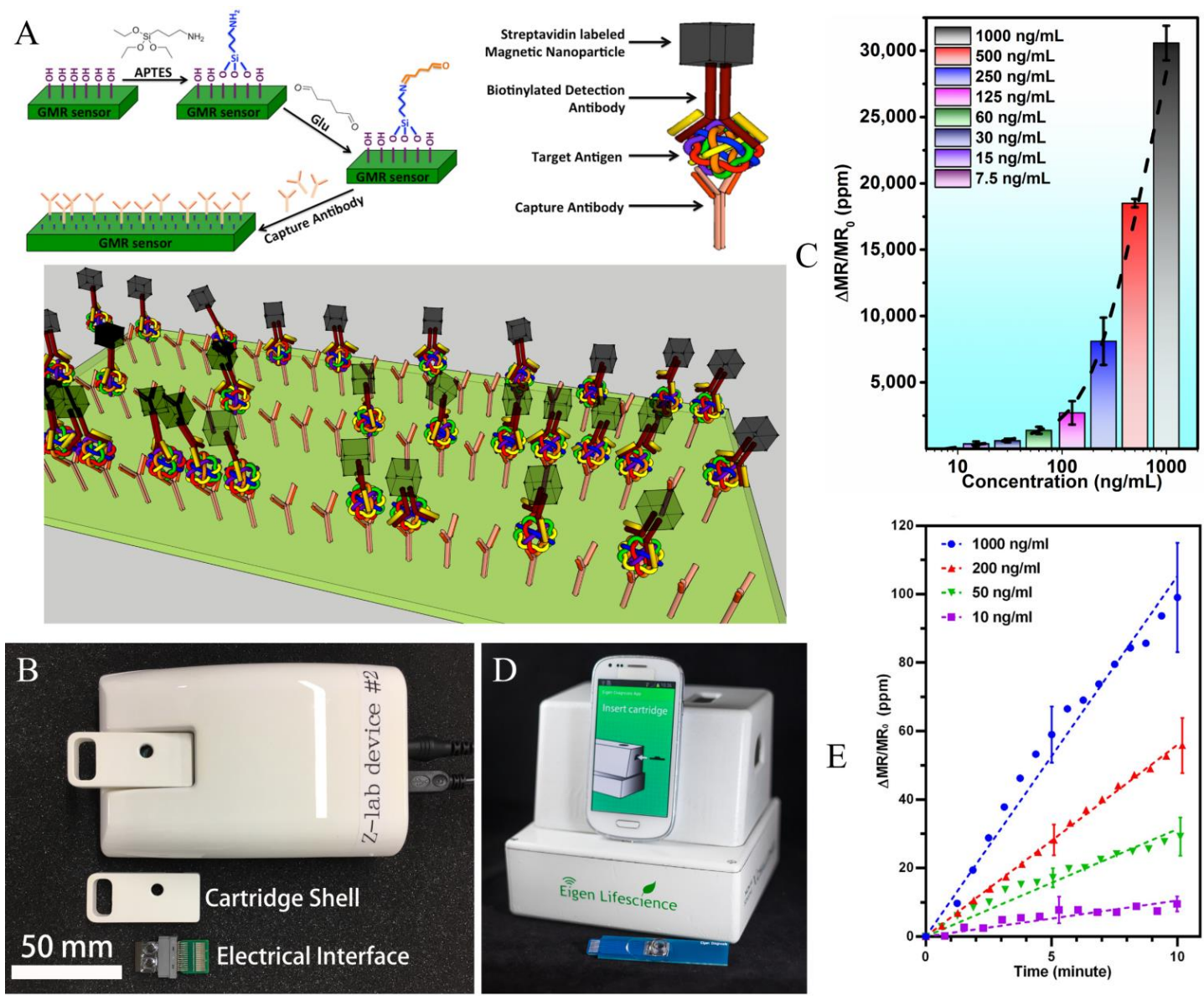

Figure 3. (A) The sandwich bioassay mechanism of a GMR biosensor forming capture antibody - target antigen - detection antibody - MNP complex. (B) Photograph of the GMR-based handheld device reported by researchers from University of Minnesota. (C) The response curves of H1N1 nucleoprotein as detected by the handheld device in (B) showing a LOD of 15 ng/mL. (D) Photograph of another GMR-based portable device reported by the researchers from Stanford University. (E) The response curves of IgG antibodies detected by device shown in (D) depicting a LOD of 10 ng/mL. (A) reprinted from [63] under the terms of the Creative Commons Attribution License (CC BY). (B) & (C) reprinted from [64], Copyright (2017) American Chemical Society. (D) & (E) reprinted from [65], Copyright (2016) Elsevier.

## 2.3 Magnetic Tunnel Junction (MTJ) Platform

The first ever proof-of-concept MTJ as biosensor was reported by Grancharov *et al*. in 2005.[70] They demonstrated a unique method for antigen and DNA detection at room temperature using monodispersed manganese ferrite nanoparticles as the magnetic tags. Since then, there have been several attempts to employ MTJs as biosensors.[61,71–73] However, most of their attempts were limited to genotyping applications of TMR sensors. In the year of 2017, Sharma *et al.* demonstrated a Poly(methyl methacrylate) (PMMA) microfluidic integrated MTJ platform (see Figure 4 (A) & (B)) for detecting pathogenic DNA from *Hepatitis E virus* (HEV), *Listeria monocytogenes*, and *Salmonella typhimurium* bacteria.[74] Figure 4(C) & (D) shows the normalized signal acquires as a function of time from MTJ sensors functionalized with HEV and *Listeria* target DNA probes, respectively, with an assay time of around 100 min. The excellent sensitivity and specificity of the microfluidic integrated MTJ platform could pave the way for lab-on-chip multiplexed apparatus and the point-of-care (POC) detection of pathogenic antigens. Very recently, Li *et al.* experimentally demonstrated the detection of HIV-1 antigen p24 by MTJ sensors with an assay time of less than 10 mins and a LOD in the orders of 0.01 µg/mL.[75]

With improved circuitry design and the ease of nanofabrication, there is a trend to use MTJ sensors for immunoassays. Gervasoni *et al*. used a 12-channel dual lock-in platform to improve the circuitry for the signal generation and acquisition in their MTJ sensing system (see Figure 4(E))[73] By customizing the differential amplifier, low-noise voltage references, and detailed analysis of temperature fluctuation within the system, they achieved a sub-ppm resolution of the lock-in amplifier and an order of magnitude better than commercial state-of-the-art instrument. However, there are several disadvantages of MTJs as biosensors compared to GMR sensors. The requirement for top electrodes increases the distance between the MNPs bound to the surface and the free layer of the MTJ sensor. As the stray fields of the MNPs decay rapidly with the increase of the distance, the sensitivity of the MTJ sensors is often sacrificed despite their high TMR ratio. Furthermore, the difficulty to achieve high linearity and low coercivity also remains a challenge for MTJs. More dedicate design of the stack structure and the fabrication process are needed to take full advantage of the high signal level induced by the large TMR ratio.

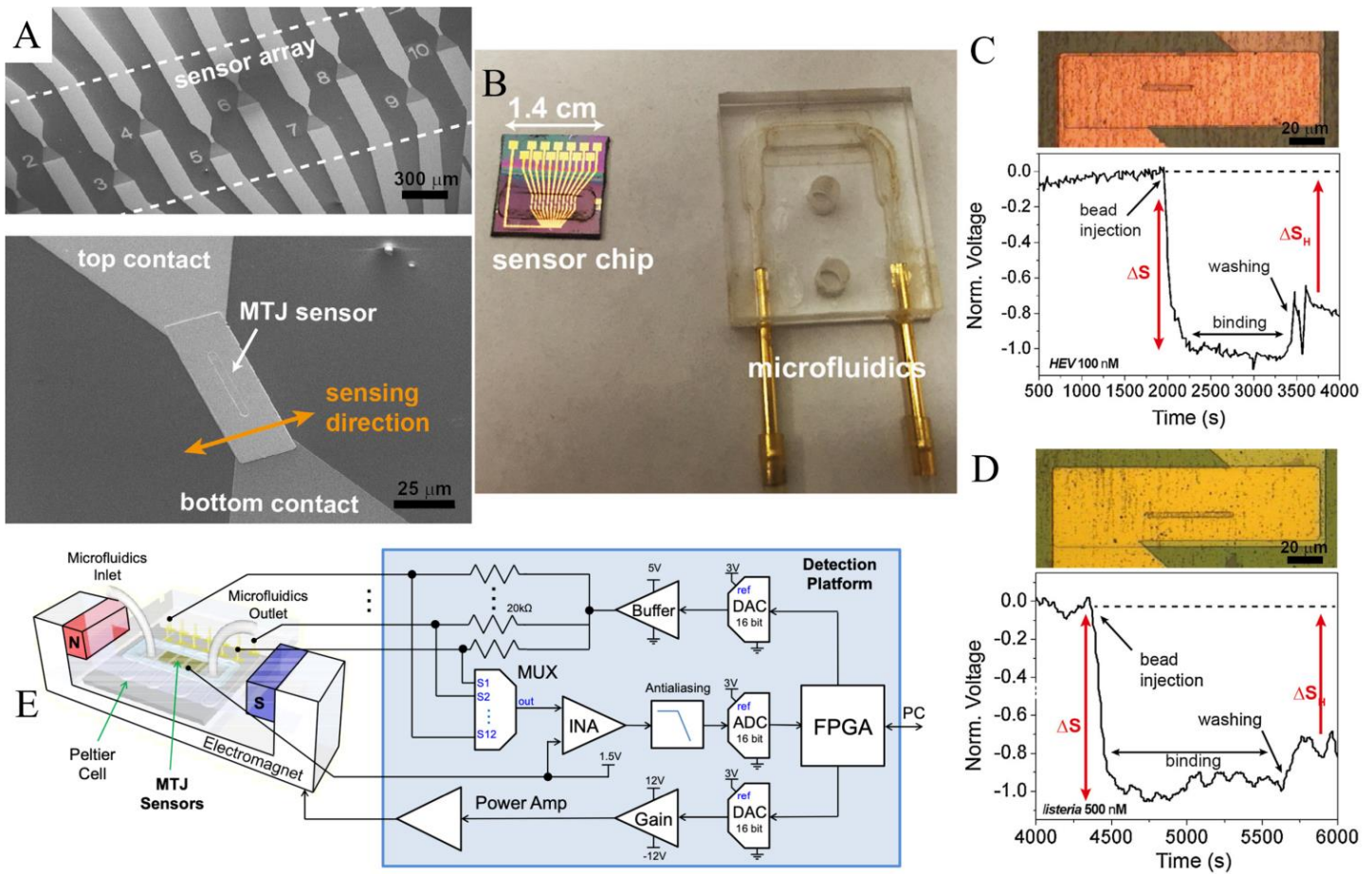

Figure 4. (A) SEM image of the sensor array comprising 12 MTJs with a common ground contact. (B) Photograph of the microfluidic channel integrated with the MTJ biosensors to facilitate handling of extremely small sample volumes. (C) & (D) are the normalized MTJ signals acquired as a function of time from sensors functionalized with (C) HEV DNA probes for detecting 100 nM HEV target DNA and (D) *Listeria* DNA probes for detecting 500 nM *Listeria* target DNA, respectively. Top panels show the photographs of sensor areas after magnetic bead immobilization. (E) The circuit schematic of the MTJ-based detection platform that mainly consists of two separate sinusoidal generators (DACs and amplifiers) and the readout circuit for 12 MTJ sensors (MUX, INA and ADC). Both the generation and processing of the input and acquired signals are performed on the FPGA. (A) - (D) reprinted with permission from [74], Copyright (2017) Elsevier. (E) reprinted with permission from [73], Copyright (2014) IEEE.

**Table 1. Comparison of Different Magnetic Immunoassay Platforms**

| Platform | Assay Time | Pathogens | Limit of Detection (LOD) | Evaluated Matrices | Ref. |
|---|---|---|---|---|---|
| GMR | < 10 mins | H1N1 | 15 ng/mL for H1N1 nucleoprotein | PBS* | [64] |
|  |  | H3N2v | 125 TCID$_{50}$/mL |  |  |
|  | < 10 mins | H1N1 | 250 TCID$_{50}$/mL | Nasal swab | [63] |
|  |  | H3N2v | 250 TCID$_{50}$/mL |  |  |

| | 15 mins | HBV | 200 IU/mL DNA | Serum | [66] |
| --- | --- | --- | --- | --- | --- |
| | N.A. | *E. coli O157H:H7* | 100 CFU/mL antigen in 1 mL sample | PBS | [68] |
| | N.A. | *Mycobacterium tuberculosis* | 1 pM ESAT-6 protein | N.A. | [69] |
| MTJ | 100 min | HEV | N.A. | PBS | [74] |
| | | *Listeria* | N.A. | | |
| | N.A. | HIV | 0.01 µg/mL antigen p24 | N.A. | [75] |
| MPS | 30 min | *Clostridium botulinum* | 0.22, 0.11, and 0.32 ng/mL for BoNT-A, -B, and -E, respectively | Milk, apple, and orange juices | [76] |
| | 25 min | *Staphylococcus aureus* | 4 and 10 pg/mL for TSST and SEA | Milk | [77] |
| | 2 h | | 0.1 and 0.3 ng/mL for TSST and SEA | | |
| | 10 s | H1N1 | 4.4 pmole for H1N1 nucleoprotein | PBS | [78] |
| | 42 min | SARS-CoV-2 | 2.96 ng/mL for SARS-CoV-2 anti-spike-protein antibodies | PBS | [79] |
| | | | 3.36 ng/mL SARS-CoV-2 anti-spike-protein antibodies | Serum | |
| NMR | 1 min | *E. coli O157:H7* | 76 CFU/mL | Water | [80] |
| | | | 92 CFU/mL | Milk | |
| | 2.5 h | *Mycobacterium tuberculosis* | 1 nM ssDNA in 1 µL sample | Sputum | [81] |

*PBS: Phosphate buffer saline

## 3. Magnetic Particle Spectroscopy (MPS) Platforms

### 3.1 Magnetic Particle Spectroscopy (MPS)

Magnetic particle spectroscopy (MPS) is firstly reported by Krause *et al.* and Nikitin *et al.* in 2006.[82,83] It is a derivative technology from magnetic particle imaging (MPI) where the tomographic images can be reconstructed by exploiting the nonlinear magnetic responses of MNPs.[84,85] Herein, in MPS-based immunoassays, the nonlinear magnetic responses of MNPs along with their rotational degree of freedom are used as metrics for different biosensing purposes. In a MPS platform, external sinusoidal magnetic fields (also called excitation fields) are applied to periodically magnetize (and magnetically saturate) the MNPs, as shown in Figure 5(A1 & A3).[76–78,86–90] The time-varying dipolar magnetic fields generated by MNPs as a response to the applied fields (see Figure 5(A2 & A4)) are monitored by pick-up coils. As a result of Faraday's law of induction, the time-varying electric voltage from pick-up coils are recorded and magnetic particle spectra are extracted for analysis, as shown in Figure 5(A3 & A6). Nowadays, there are two excitation field modes of MPS platforms that have been frequently reported: the mono-frequency and dual-frequency modes. In a mono-frequency MPS platform, one sinusoidal magnetic field with frequency f is applied and higher odd harmonics at 3f (the 3$^{rd}$ harmonic), 5f (the 5$^{th}$ harmonic), 7f (the 7$^{th}$ harmonic), … are observed due to the nonlinear magnetic responses of MNPs.[86,87,91]

On the other hand, in a dual-frequency MPS platform, two sinusoidal magnetic fields with frequencies $f_H$ and $f_L$ are applied. The low frequency field $f_L$ periodically magnetizes MNPs while the high frequency field $f_H$ modulates these higher odd harmonics to high frequency range. Thus, higher odd harmonics at $f_H \pm 2f_L$ (the 3$^{rd}$ harmonics), $f_H \pm 4f_L$ (the 5$^{th}$ harmonics), $f_H \pm 6f_L$ (the 7$^{th}$ harmonics), … are observed.[88,90,92–95] Although different in excitation modes, the detection mechanisms of periodically magnetize the MNPs and the extraction of higher odd harmonics as a result of nonlinear magnetic responses are identical.

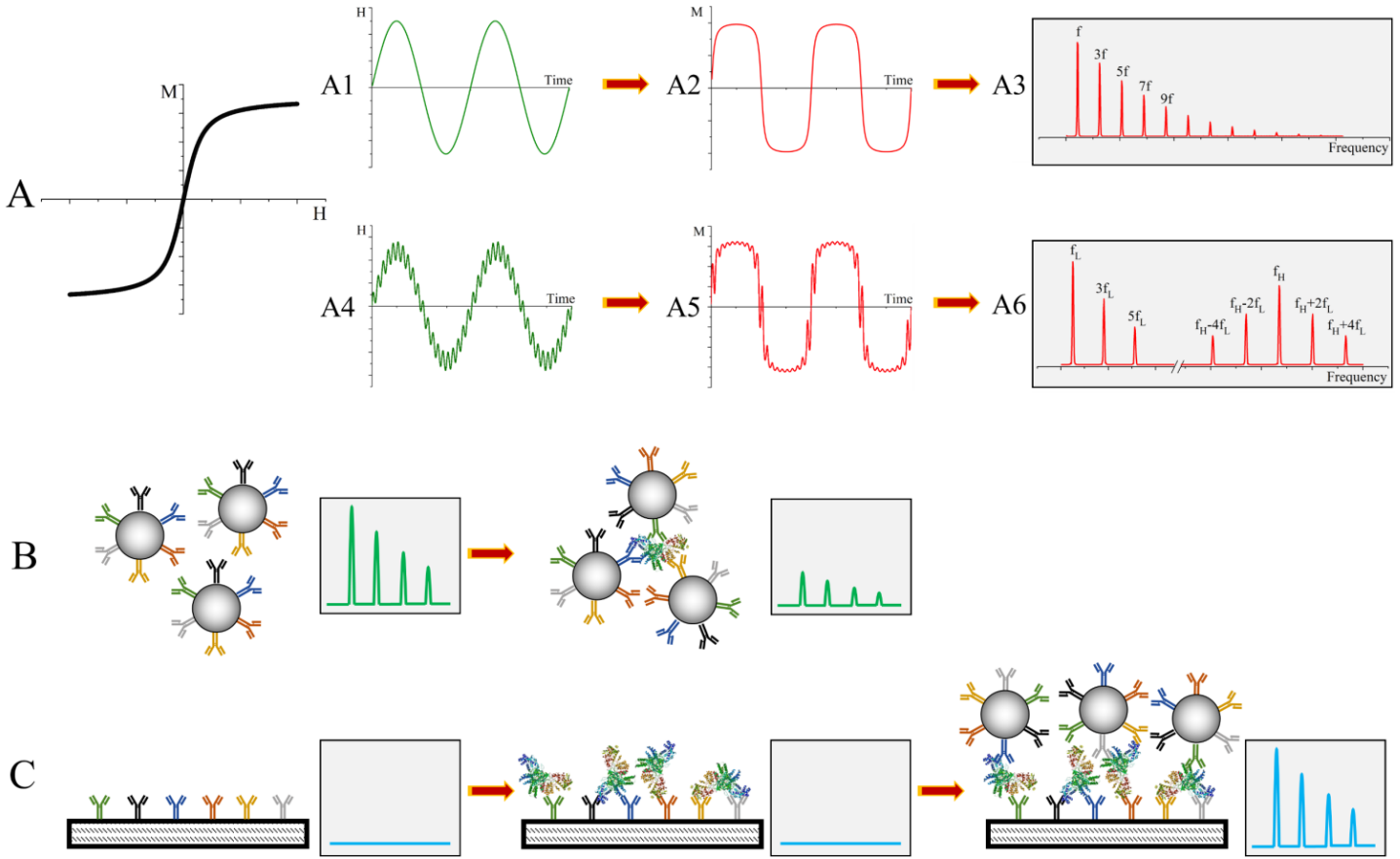

Figure 5. (A) Nonlinear magnetic responses of MNPs. (A1-A3) and (A4-A6) are the mono- and dual-frequency modes. A1 & A4 are the time domain excitation fields. A2 & A5 are the time domain magnetic responses. A3 & A6 are the magnetic particle spectra extracted from the pick-up coils. (B) Schematic drawing of the volume-based MPS immunoassay. (C) Schematic drawing of the surface-based MPS immunoassay.

In addition, there are two types of MPS-based immunoassay platforms: the volume- and surface-based platforms (see Figure 5(B&C)). Although both platforms use dynamic magnetic responses of MNPs for characterizations, the degrees of freedom are different. In volume-based MPS platform, MNPs are dispersed in liquid phase. Upon the application of external magnetic fields, their magnetic moments relax to align to the external fields through the joint Brownian and Néel relaxation processes. Where Brownian relaxation is the physical rotation of whole MNP with its fixed magnetic moment and Néel relaxation is the rotation of the magnetic moment inside a stational MNP. For volume-based MPS platforms, single-core, superparamagnetic nanoparticles (SPMNPs) that realign magnetic moments to external fields through a Brownian relaxation-dominated process are favored. The Brownian relaxation process is affected by liquid viscosity, hydrodynamic volume of MNP, and temperature (note: other factors such as magnetic field amplitude, dipolar interactions between neighboring MNPs, magnetic properties of MNPs such as saturation magnetization, anisotropy, etc. are not in the scope of this review).[94–100] By surface functioning MNPs with biological/chemical reagents such as antibodies, DNA, RNA, proteins, the MNPs serve as high specificity probes to capture target analytes from biofluid samples. As shown in Figure 5(B), the successful recognition and binding events on MNPs cause increased hydrodynamic volume. Thus, Brownian relaxation is blocked, and magnetic responses are weakened. Larger phase lags between magnetic moments and the external fields are detected and lower harmonic amplitudes are observed from the magnetic particle spectra. In this volume-based MPS platform, the immunoassay is achieved by monitoring the reduced rotational freedom of MNPs in the testing suspension. On the other hand, in surface-based MPS platform, surface functionalized MNPs are captured to a solid substrate (i.e., reaction surface) and their Brownian rotational freedom are blocked, as shown in Figure 5(C). As a result, immunoassays are achieved by "counting" the number of MNPs captured to the solid substrate. In this section, different MPS platforms for virus and pathogen detections are reviewed and comparisons are made in Table 1.

*3.2 Surface-based MPS Platform*

Orlov *et al.* reported a multiplexed lateral flow (LF) assay for on-site detection of botulinum neurotoxin (BoNT) types A, B, and E from complex matrixes.[76] The BoNT-A, -B, and -E are proteins produced by anaerobic bacteria of *Clostridium botulinum* widely present in soil and water. In their work, the multiplexing is realized by combing MPS platform with LF measurement. LF method is based on various optical labels such as latex, gold, silver, and quantum dots, this method alone is difficult to achieve high sensitivity, quantitative immunoassays, especially in opaque mediums.[101–105] By replacing these optical labels with magnetic labels (i.e., MNPs), a potentially high sensitivity, high stability and low background noise (regardless of the optical transparency of biological media) biosensing platform is achieved. Herein, the authors combined three single-plex test strips with dissimilar positions of the test lines in a miniature cartridge. Each test strip is named A-strip, B-strip, and E-strip, respectively, for intended detection of BoNT-A, B, and E, respectively, as shown in Figure 6(D). Each strip is composed of overlapping sample pad, conjugation pad, nitrocellulose, and wicking pad on an adhesive plastic backing sheet, as show in Figure 6(A). The anti-BoNT capture antibodies (labeled as capture Ab in the figure) are deposited onto the nitrocellulose membrane as labeled test line. The corresponding MNP-detection antibody complexes (labeled as MP-Ab in the figure) are deposited on the conjugation pad. During an assay process, testing sample is deposited onto the sample pad and the fluid migrates along the test strip under the capillary action. The target analytes bind to MP-Ab and capture Ab on the test line. As shown in Figure 6(B), the distributions of MNPs along the test strip exhibit three peaks corresponding to: remaining MNPs left in conjugation pad, MNPs bind on test line due to the presence of target analytes, unbounded MNPs collected at the wicking pad. The magnetic signal amplitudes recorded by MPS (labeled as MPQ reader in the figure) are positively correlated with the concentration (quantity) of target analytes. The specifically captured MNPs on the test line of the nitrocellulose membrane can be seen in the scanning electron microscope (SEM) image shown in Figure 6(C). The multiplexed assay procedures and measuring setup are like the single-plex assay, by replacing the single-plex strip with a cartridge. Sample is deposited on the front end of cartridge and after ~25 min the cartridge is inserted into the MPQ reader for measurements. Using this method, the authors have successfully and simultaneously detected three botulinum toxin serotypes from complex liquid food matrixes such as whole milk and juices.

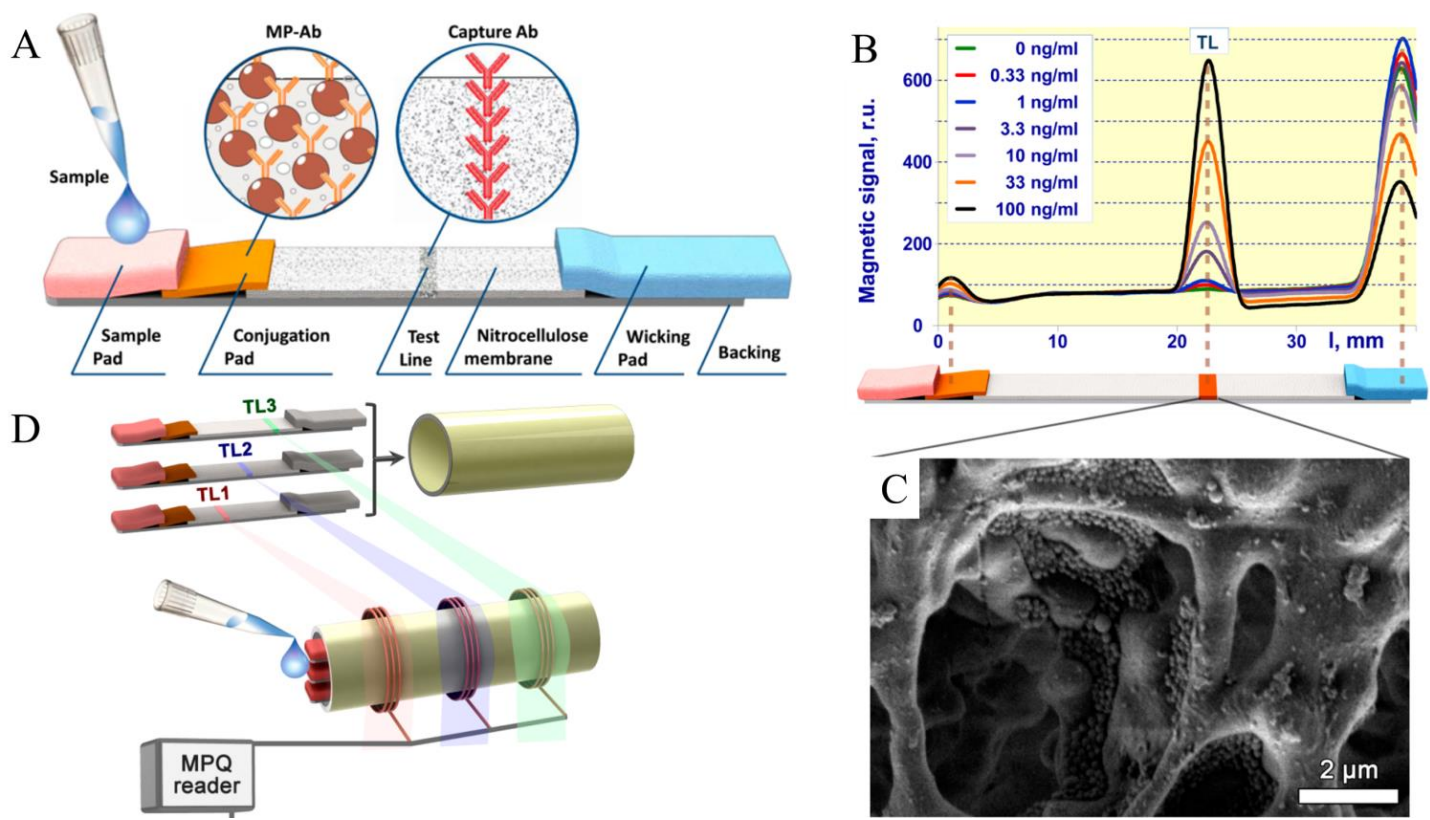

Figure 6. (A) Test strip design based on sandwich-lateral flow assay with antibody-conjugated MNPs as labels. (B) Typical distributions of MNPs along the lateral flow test strip for different concentrations of BoNT-A. (C) SEM image of MNPs specifically captured at the test line on the membrane. (D) Multiplexed assay setup: several single-plex test strips with dissimilar positions of the test lines are combined in a miniature cartridge. The cartridge with a sample deposited onto its front end is inserted into the portable magnetic particle quantification (MPQ) reader (i.e., MPS). Figure reprinted with permission from [76], Copyright (2016) American Chemical Society.

The authors successfully combined MPS method with lateral flow method. By conjugating different capture antibodies onto different locations of a test strip, a multiplexed assay platform is achieved. By replacing the optical labels with MNPs, the measurements can analyze media regardless of the optical properties, offering sensitivity on the level of lab-based quantitative methods.

Orlov *et al.* reported the application of MPS platform for detection of toxins produced by *Staphylococcus aureus*.[77] Due to the high stability and increasing resistance to antibacterial medications of these bacteria, the toxins are widely present in the environment and are frequently responsible for diverse fatal illnesses such as severe gastrointestinal diseases and toxic shock. In their work, they introduced a novel magnetic immunoassay on 3D fiber solid phase (see Figure 7(D)) that fits into a standard automatic pipet tip, as shown in Figure 7(A). The 3D porous filter surfaces are immobilized with capture antibodies specific to a definite toxin. These as-prepared solid-phase filter immobilized with antibodies can be either used immediately or stored for a long time without compromising the properties. Two measurement formats are proposed: one for analysis of small volume samples (Figure 7(B), labeled as express MIA) and the other for large volume samples (Figure 7(C), labeled as high-volume MIA). In the express MIA, samples are dispensed simultaneously through all the tips by an electronic pipet. In the high-volume MIA, testing sample is pumped through the 3D fiber filters and the sample volume is determined by the pumping rate and time. In this step, the target analytes flowing through 3D porous filters are captured by the capture antibodies from the solid-phase filter. Further steps are the same for both formats. After passing the samples through the filters, each filter is washed to removed unbound reagents. Then 7-min dispensing of detection antibody-MNP complexes are carried out followed by another cycle of washing step. The MNPs that bound to the immunocomplex on the 3D porous fiber surfaces serve as labels to be recorded by the MPS reader.

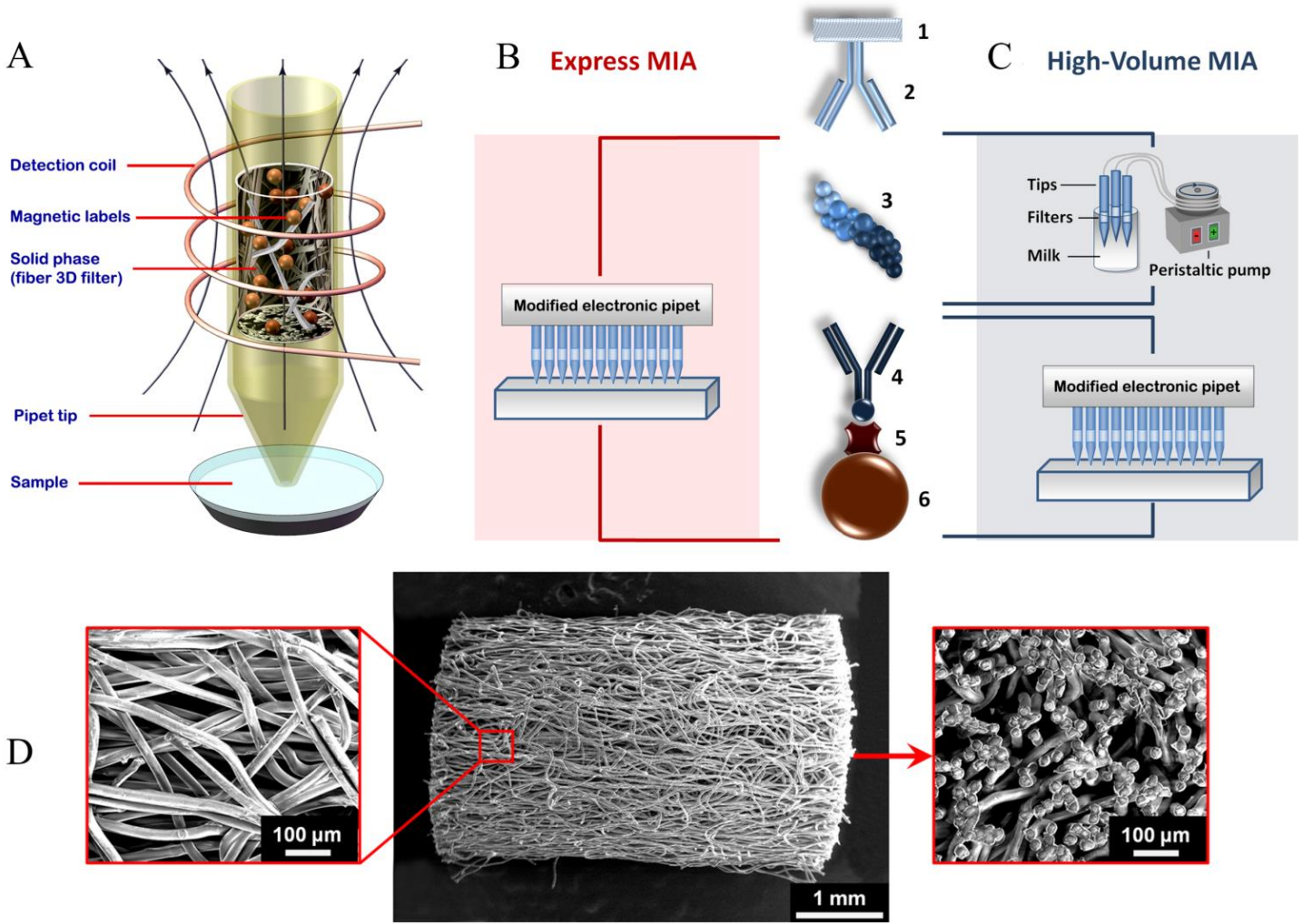

Figure 7. (A) Schematic drawing of 3D porous filters as a solid phase immunoassay substrate in a cylinder. (B) & (C) are the schematic drawing of sandwich structure magnetic immunoassay on 3D fiber filters. 1, filter surface; 2, capture antibody; 3, antigen; 4, biotinylated detection antibody; 5, streptavidin; 6, MNP. (A) Express MIA setup and (B) high volume MIA setup. (D) SEM surface morphology of cylindrical 3D fiber filters. Left, magnified lateral surface fragment (500× magnification); center, overview of lateral surface of the filter (60× magnification); right, top surface fragment (500× magnification). Figure reprinted with permission from [77], Copyright (2012) American Chemical Society.

In June 2020, Pietschmann *et al.* reported the portable MPS surface-based immunoassay platform, MInD (Magnetic Immuno-Detection) for the detection of SARS-CoV-2 specific antibodies.[79] In their work, a porous polyethylene filter matrix coated with SARS-CoV-2 spike protein peptide is serving as the reaction surface (called immunofiltration columns in the paper). Varying concentrations of SARS-CoV-2 anti-spike-protein antibodies in PBS and human serum samples are spiked through the surface, followed by a wash step to remove unbound antibodies. Then biotinylated secondary antibodies are added and followed by another wash step. Finally, streptavidin-coated MNPs are added to the reaction surface, forming a [SARS-CoV-2 spike protein peptide] – [SARS-CoV-2 anti-spike-protein antibody] – [secondary antibody] – [MNP] structure. After a final wash step, the MPS spectra of captured MNPs are measured. They achieved LODs of 2.96 ng/mL and 3.36 ng/mL for detecting SARS-CoV-2 anti-spike-protein antibodies from PBS and human serum, respectively. It shows better sensitivity and wider detection range than commonly used analytical biochemistry assay ELISA (enzyme-linked immunosorbent assay). However, negative control groups are PBS and serum without anti-spike-protein antibodies. The detection of antibodies generated in response to the infection can provide a larger window of time for indirectly detecting SARS-CoV-2. Antibody testing is very useful for the surveillance of COVID-19. One potential challenge of developing accurate antibody detections is the potential cross-reactivity of SARS-CoV-2 antibodies with antibodies generated against other coronaviruses.[26] Yet in this work, the cross-reactivity with another coronavirus such as MERS coronavirus (MERS-CoV) and SARS coronavirus (SARS-CoV) are not tested. This point-of-care (POC) testing device allows for identifying people with immunity against SARS-CoV-2.

### 3.3 Volume-based MPS Platform

Wu *et al.* reported the volume-based MPS immunoassay platform utilizing the polyclonal antibodies induced cross-linking of MNPs for one-step, wash-free detection of H1N1 nucleoprotein molecules.[78] In their work, the MNPs are anchored with polyclonal IgG antibodies specific to H1N1 nucleoprotein. The H1N1 nucleoprotein molecule hosts multiple epitopes that serve as binding sites for IgG polyclonal antibodies. Thus, each nucleoprotein can bind to more than one MNPs, consequently assembling into MNP clusters. As shown in Figure 8(A), seven experimental groups and two negative groups are prepared. For negative control 2 (sample index IX), magnetic responses of bare MNPs are recorded in the MPS platform. For negative control 1 (sample index VIII), the magnetic responses of polyclonal antibody-MNP complexes are recorded. For experimental groups I – VII, different concentrations of H1N1 nucleoprotein are mixed with polyclonal antibody-MNP complexes, the concentrations from highest to lowest are 4.42 μM (I), 2.21 μM (II), 884 nM (III), 442 nM (IV), 221 nM (V), 88 nM (VI), and 44 nM (VII). Due to the varying abundancies of target analytes (i.e., H1N1 nucleoprotein), different degrees of MNP clustering are observed from samples I - VII. As shown in Figure 8(B), with the increasing degree of MNP clustering, the averaged MNP hydrodynamic size increases and the harmonic amplitude decreases. Figure 8(C) shows the $3^{rd}$ and the $5^{th}$ harmonic amplitudes from samples IX, VIII, and I. With the anchoring of polyclonal antibodies onto MNPs, a small decrease in harmonic amplitude is observed from sample VIII compared to sample IX. Which proves the successful conjugation of antibodies on MNPs and as a result, the hydrodynamic size slightly increases. The experimental group (sample I) shows substantial decrease in harmonic amplitudes due to the H1N1 nucleoprotein induced MNP clustering. As a side note, the harmonic ratios are also used as MNP-quantity independent factor for MPS-based immunoassays. Figure 8(D: [a] – [e]) shows the hydrodynamic size distributions of MNPs from samples II (2.21 μM), IV (442 nM), VI (88 nM), VIII (MNP-antibody complex), and IX (bare MNP) measured by dynamic light scatter (DLS). The hydrodynamic size increases after the anchoring of antibodies onto MNPs, then further increases in the presence of H1N1 nucleoprotein. Figure 8(D: [f]) gives a more intuitive comparison between samples II (2.21 μM) and VIII (MNP-antibody complex). The H1N1 nucleoprotein causes noticeable size peak shift from 46 nm to 59 nm. In addition, the bump between 200 nm and 300 nm indicates the presence of MNP clusters. In Figure WK4(E) & (F), the harmonic amplitudes recorded from samples I-IX show the similar trends.

This one-step, wash-free, volume-based MPS detection scheme allows for immunoassays on minimally processed biological samples and handled by non-technicians with minimum training requirements. Since the magnetic signals come for the whole volume of MNP suspension thus, removing the unbound MNPs from the sample could ensure higher detection sensitivity for this type of volume-based assay mode.

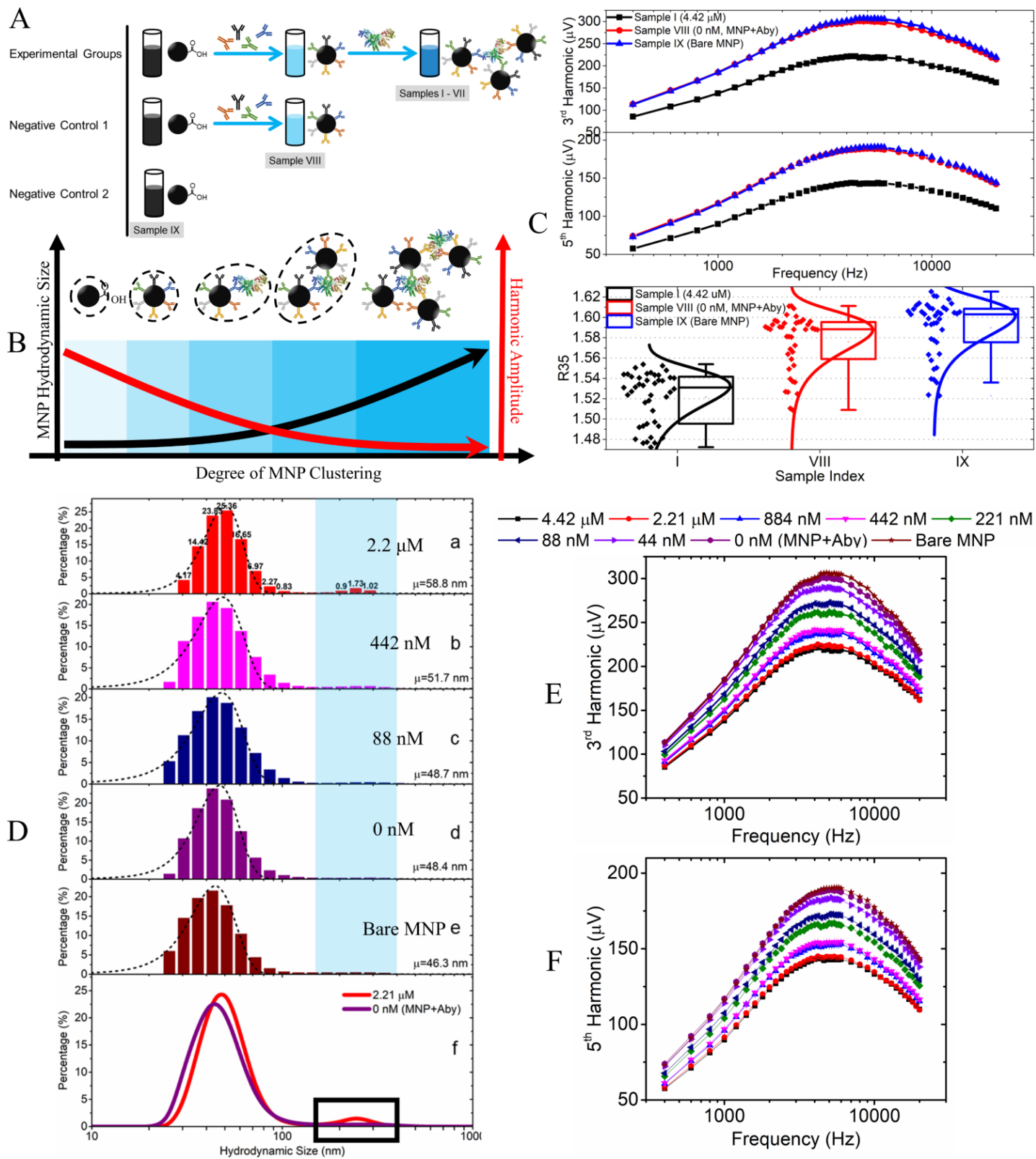

Figure 8. (A) Sample preparation flowcharts of the experimental and negative control groups. Nine MNP samples are prepared: sample indexes I−VII are MNP-antibody complexes in the presence of different concentrations of the H1N1 nucleoprotein; sample index VIII is an MNP-antibody complex in the absence of the H1N1 nucleoprotein (denoted as "0 nM (MNP+Aby)"); sample index IX is bare MNP suspension (denoted as "Bare

MNP"). (B) The 3$^{rd}$ and the 5$^{th}$ harmonics along varying driving field frequencies (only samples I, VIII, and IX are plotted) collected by the MPS system. Boxplots of the harmonic ratios (R35) collected from samples I, VIII, and IX. (D) Statistical distribution of the hydrodynamic sizes of samples [a] II, [b] IV, [c] VI, [d] VIII, and [e] IX as characterized by DLS. [f] Comparison of the measured DLS size distribution curves between samples II (2.21 µM) and VIII (0 nM, MNP+Aby). (E) and (F) are MPS measurements of the 3$^{rd}$ and the 5$^{th}$ harmonics from samples I−IX at varying driving field frequencies from 400 Hz to 20 kHz. Figure reprinted with permission from [78], Copyright (2020) American Chemical Society.

## 4. Nuclear Magnetic Resonance (NMR) Platforms

### *4.1 Nuclear Magnetic Resonance (NMR)*

Nuclear magnetic resonance (NMR) detects the MNP-labeled targets by measuring the precessional signal of $^1$H proton from the whole sample volume. In this way, the NMR platform is categorized as one type of volume-based immunoassay methods. Note: NMR-based immunoassay platform is also called magnetic relaxation switching (MRS). As shown in Figure 9(A), due to the high surface-to-volume ratio of MNPs, the local magnetic field inhomogeneity caused by MNP disturbs the precession frequency variations in millions of surrounding water protons, which accelerates the decay of the spin system's phase coherence. In addition, the NMR-based detection intrinsically benefits from signal amplification and is able to achieve high detection sensitivity. As the mono dispersed MNPs aggregates upon binding to targets, the self-assembled clusters become more efficient in dephasing the nuclear spins of surrounding water protons, resulting in decreased $T_2$ relaxation time. The reverse is also true upon the cluster disassembly. Figure 9(B) shows the steps of NMR-based immunoassay with MNP-pathogen interaction, magnetic separation, and filtration. As is mentioned in Section 3.3, for volume-based biosensing platforms, the filtration step could effectively reduce the interference of unbound MNPs. The magnetic separation and filtration are not necessary but favored for high sensitivity immunoassays.

Issadore *et al.* reported a miniaturized NMR platform for point-of-care (POC) diagnostics.[106] A photograph and the schematic of the portable NMR platform are shown in Figure 9(C) & (D). The portable magnet, microcoil, and RF (radio frequency) matching circuit are packaged into a thermally insulating PMMA (poly-methylmethacrylate) housing. The custom electronics provides the NMR pulse sequences, collects the NMR signal, and communicates with external terminals. Samples are loaded into thin-wall polyimide tubes and introduced into the coil bore for NMR measurements. The modular coils can be plugged into the system to optimally accommodate sample volumes from 1 mL to 100 mL. This portable NMR platform with automatic measurement setting tuning provides users with easy-to-use interface and offers sensitive on-site diagnosis. With these capabilities, it is expected that NMR handheld device can be an essential tool for personal care and accurate diagnostics for infectious diseases in resource-limited areas, mitigating the burden in public health.

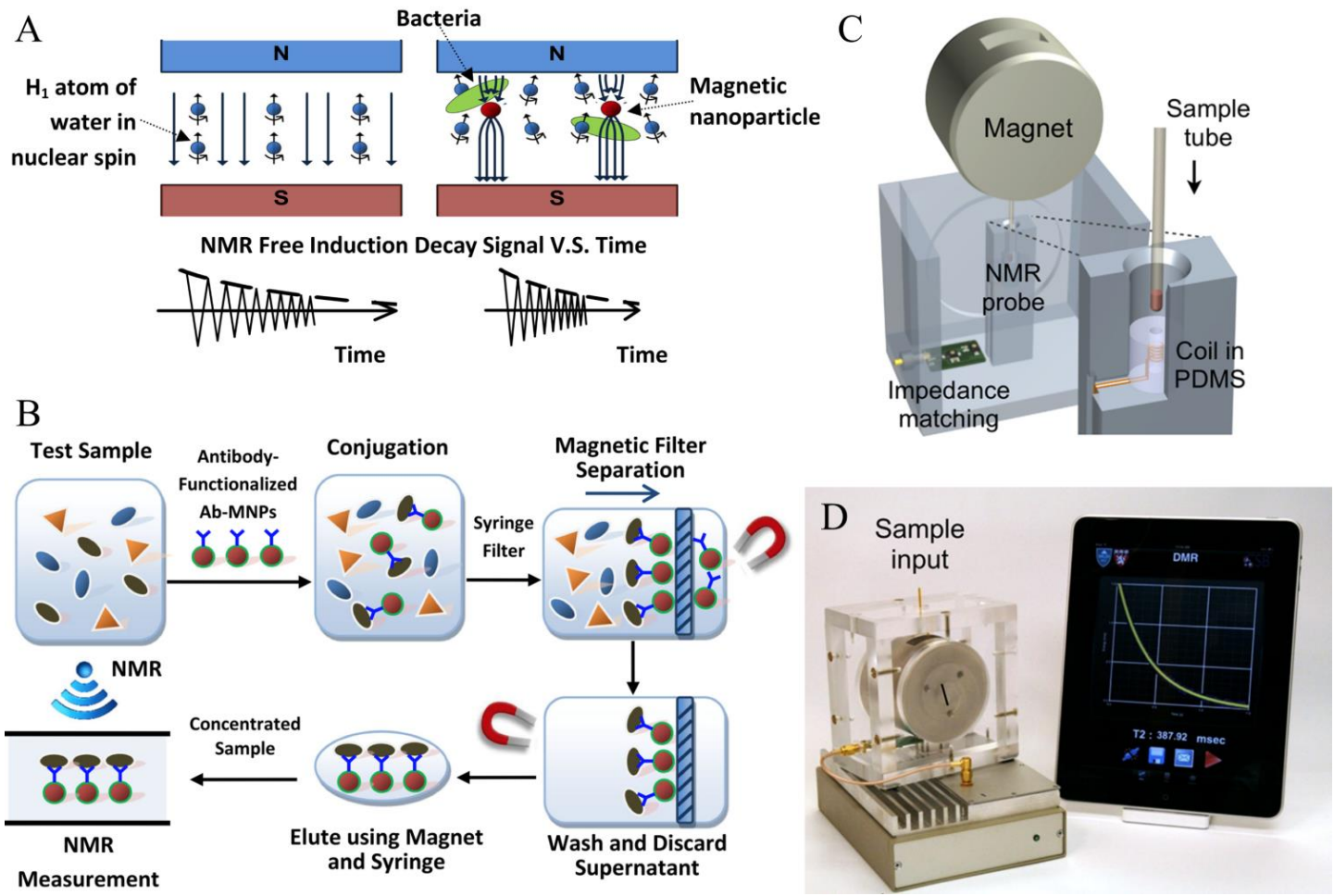

Figure 9. (A) The MNP induced spatial and temporal disturbance in the homogeneity and strength of the local magnetic field. (B) Schematic drawing of the working principle of the NMR based biosensor for pathogen detection. A filtration step is included in this scheme, thus, the interference of unbound MNPs can be effectively reduced, and MNPs in the eluted solution are proportional to the pathogen concentration. (D) Photograph of a portable NMR device. This system has a capacity for automatic system tuning and features a user-friendly interface. (D) Magnet assembly and the NMR probe design. The microcoil is embedded in a polydimethylsiloxane (PDMS) substrate. The entire coil bore is accessible by a sample, which maximizes the sample-filling factor. A thin-walled tube is used for sample loading. (A) & (B) are reprinted from [80] under the terms of the Creative Commons Attribution 4.0 International License. (C) & (D) are reprinted with permission from [106], Copyright (2011) The Royal Society of Chemistry.

## *4.2 NMR-based Immunoassays*

Recent advances in micro- and nanofabrication has accelerated the development of portable NMR devices. Luo *et al.* reported the detection of foodborne bacteria *Escherichia coli O157:H7* from drinking water and milk samples using a portable NMR platform.[80] The NMR system is able to generate 0.47 T of magnetic field and a high-power pulsed RF transmitter with ultra-low noise sensing circuitry capable of detecting weak NMR signal at 0.1 µV. In their work, the bacteria are labeled with MNPs through the antibody-pathogen interactions. A 20 – 30 min filtration step is carried out and followed by 1 min of NMR signal collection.

Liong *et al.* reported the detection of nucleic acids based on a magnetic barcoding strategy.[81] Where the PCR-amplified mycobacterial genes are sequence-specifically captured on microspheres, labeled by MNPs, and detected by NMR. All the components and steps are integrated into a single, small fluidic cartridge for streamlined on-chip assays. As shown in Figure 10(A), the sputum samples are first processed off-chip to extract DNA from *mycobacterium tuberculosis* and followed by PCR amplification. The amplicons are captured by polymeric beads modified with complementary capture DNA strands. Then MNPs modified with probed DNA strands bind to the opposite end of the amplicon. This capture DNA – target DNA – probe DNA scheme enhances the detection specificity and offers fast binding kinetics. After the removal of unbound MNPs, samples are subjected to NMR measurements. The MNPs captured due to target DNA cause faster relaxation of the $^1$H NMR signal and the decay rate is proportional of the MNP amount (and the amount of initial DNA), enabling the quantification of target DNA strands. The microfluid device for on-chip NMR measurements is shown in Figure 10(B). MNPs and buffers are preloaded in gated chambers, after the target DNA strands are PCR-amplified, the amplicons are combined with capture beads. The bead-DNA complexes are then mixed with MNPs and introduced into the mixing channel. The MNP labeled beads are filtered by an in-line membrane shown in Figure 10(C) and concentrated into the NMR chamber for measurements.

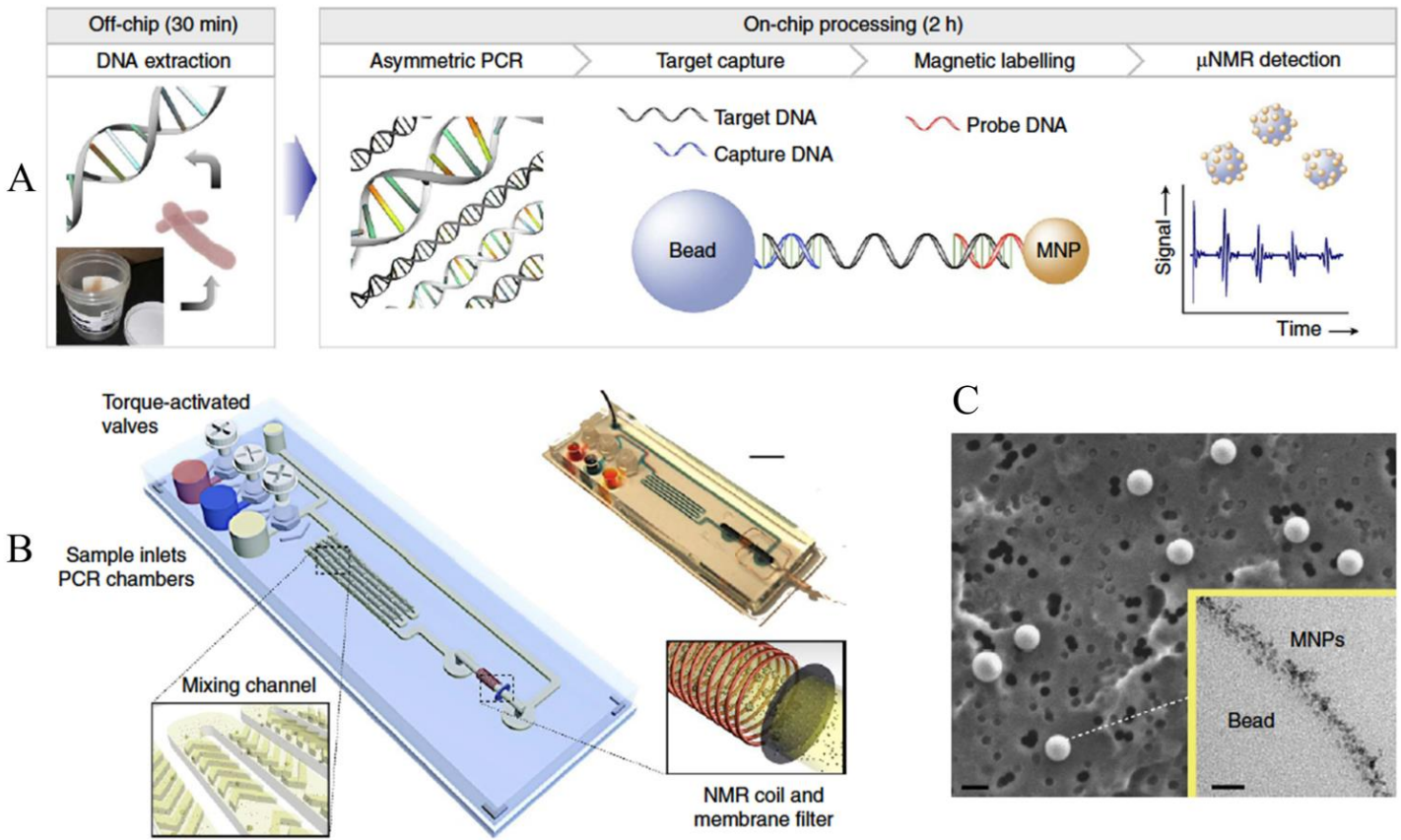

Figure 10. (A). Assay procedure. (B) Fluidic cartridge integrating PCR chambers, mixing channel, and a microcoil for NMR measurements. The entire cartridge is disposable to prevent cross-contamination of PCR-amplified products. (C) SEM image of the bead captured by the membrane filter, scale bar, 1 μm. The inset SEM image shows that the beads are efficiently labelled with MNPs, scale bar, 30 nm. Figure reprinted with permission from [81], Copyright (2013) Macmillan Publishers Limited.

In addition to the immunoassay applications, saturation transfer difference (STD) NMR has emerged as a robust tool for characterizing protein binding and ligand screening.[107] It is used for identifying the underlying mechanisms of Hepatitis B virus X protein (HBx)-mediated carcinogenesis. Yue *et al.* used NMR-based metabolomics approach to systematically investigate the effects of HBx on cell metabolism.[108] Kusunoki *et al.* used NMR to characterize the interactions between the HBx BH3-like motif and Bcl-$_{XL}$ and showed that this motif binds to the common BH3-binding hydrophobic groove of Bcl-$_{XL}$ with a binding affinity of 89 μM.[109] NMR is applied for assessing the ability of an artificially designed oligopeptide in binding to Ebola virus Viral Protein 24 (VP24).[110] The successful protein-protein binding could inhibit the Ebola virus VP24 protein's interaction with the human protein Karyopherin, thus, reduce the Ebola virus virulence. Vasile *et al.* used NMR to study the interactions between the sialic acid and influenza hemagglutinin (HA) from human and avian strains.[111] By screening the HA ligand-protein interactions, it could yield useful information for an efficient drug design.

Herein, we have reviewed different magnetic immunoassay platforms and included the most representative literatures. The advantages and disadvantages of each platform are listed and compared in Table 2.

**Table 2. Advantages and Disadvantages of Different Magnetic Immunoassay Platforms**

| Platforms | Advantages | Disadvantages |
|---|---|---|
| GMR | - High sensitivity.<br>- Portable device available.<br>- Mass production capability | - Requires multiple wash steps thus, well trained technicians are needed.<br>- Time consuming.<br>- High cost per test, nanofabrication of GMR biosensors is required. |
| MTJ | - High sensitivity.<br>- Mass production capability | - Requires multiple wash steps thus, well trained technicians are needed.<br>- High noise. Large distance from MNP to sensor surface.<br>- Hard to acquire linear response.<br>- Complicated fabrication process.<br>- Time consuming. |
| | | • High cost per test, nanofabrication of MTJ biosensors is required. |
| MPS, surface-based | • High sensitivity. <br> • Low cost per test. <br> • Portable device available. | • Requires multiple wash steps thus, well trained technicians are needed. <br> • Time consuming. |
| MPS, volume-based | • Allows one-step wash-free detection. <br> • Immunoassays can be handheld by non-technicians. <br> • Low cost per test. <br> • Portable device available. | • Requires multiple wash steps thus, well trained technicians are needed. <br> • Medium sensitivity |
| NMR | • High sensitivity. <br> • Portable device available. | • Requires multiple wash steps thus, well trained technicians are needed. <br> • Time consuming. |

## 5. Other Magnetic Immunoassay Methods

In most magnetic immunoassay platforms, MNPs are used as labels (e.g., MR sensors and MPS platforms) or contrast enhancers (e.g., NMR platforms) due to the unique magnetic properties, large surface-to-volume ratio, good stability and biocompatibility, and the facile surface functionalization with a great variety of reagents. In addition to the above technologies, other platforms that utilize MNPs as auxiliary tools for virus and pathogen detections have also been extensively reported. In this section, we reviewed some representative works that use magnetic materials are auxiliary tools for high sensitivity virus and pathogen detections, as summarized in Table 3.

Chou *et al.* use surface functionalized MNPs as probes for efficient magnetic separation to achieve rapid and sensitive virus screening.[112] In their work, MNPs are functionalized with H5N2 viral antibodies targeting the hemagglutinin (HA) protein and capped with methoxy-terminated ethylene glycol to block nonspecific binding. Combined with magnetic separation, these MNPs show effective isolation of H5N2 from lysate for direct matrix-assisted laser desorption/ionization time-of-flight mass spectrometry (MALDI-TOF MS) readout without any elution steps. A limit of detection in the range of $10^{4.5} - 10^{5.5}$ TCID$_{50}$ is achieved within a diagnosis time of one hour. The functionalized MNP probes can unambiguously differentiate the H5N2 viruses from other closely related subtypes such as H5N1 viruses in a highly specific manner thus, can be utilized for the rapid screening of virus subtypes.

Tian *et al.* reported the ferromagnetic resonance (FMR) based homogeneous and volumetric biosensor for DNA detection.[113] This method quantifies the target DNA by measuring the FMR resonance field shift of the suspension. The detection strategy for target DNA is based on an isothermal amplification followed by hybridizing with detection antibody modified MNPs. In the presence of target DNA strands, antibody-MNP complexes form aggregates and which lowers the net anisotropy as well as increases of the resonance field. For the rolling circle amplification (RCA)-based FMR assays, a LOD of 1 pM and linear detection range of 7.8 – 250 pM are obtained for detecting synthetic Vibrio cholerae target DNA from buffer solutions. For the loop-mediated isothermal amplification (LAMP)-based FMR assays, a LOD of 100 aM is obtained for detecting synthetic Zika virus target oligonucleotide from 20% serum samples.

Sun *et al.* reported a magnetic immunoassay method based on surface enhanced Raman scattering (SERS) spectroscopy to detect influenza virus H3N2 (A/Shanghai/4084T/2012) through sandwich-structure complex consisting of: SERS tags, target influenza virus and highly SERS-active $Fe_3O_4$/Au MNPs as supporting and capturing substrates.[114] Using a portable Raman spectrometer, a LOD of $10^2$ $TCID_{50}$/mL and linear detection range of $10^2$ to $5 \times 10^3$ $TCID_{50}$/mL are achieved.

Barrios-Gumiel *et al.* reported the carbosilane dendrons decorated MNPs with peripheral carboxyl and carboxylate groups for the capture and concentrate of R5-HIV-1$_{NLAD8}$ and X4-HIV-1$_{NL4.3}$ strains.[115] The carboxyl and carboxylate MNPs assist in achieving rapid and easy diagnostics and reduce/eliminate the risk of HIV-1 transmission.

Wang *et al.* employed two kinds of labels for the virus antibody and antigen separately.[116] The virus antigens were functionalized with fluorescent encoded MNPs, while the antibodies were conjugated to green-emitting CdTe quantum dots (QDs). By applying different kinds of fluorescent nanocomposites to the antigens, multiplexed detection of equine influenza virus (EIV) and equine infectious anemia virus (EIAV) was achieved with a sensitivity of 1.3 ng/mL and 1.2 ng/mL for EIV antigens and EIAV antigens, respectively.

Zhang *et al.* synthesized a virus-magnetic-molecularly imprinted polymer (MIP) complex under an applied magnetic field.[117] The existence of $Fe_3O_4$ MNPs can accelerate the preparation process of the complex. As the viruses were captured specifically to the surface of the magnetic-MIP, the size and shape of the particles changed, leading to a change in the magnetic resonance light scattering (RLS) signal. The linear concentration range for hepatitis A virus was 0.02-1.4 nM, with a detection limit of 6.2 pM. A similar setup was also employed to detect the Japanese encephalitis virus (JEV) with a limit of detection of 1.3 pM in human serum.[118]

Ali *et al.* employed MNPs in a reverse transcription polymerase chain reaction (RT-PCR) platform for multiplexed detection of hepatitis B virus (HBV), hepatitis C virus (HCV), and human immunodeficiency virus (HIV).[119] Silica coated MNPs were used during nucleic acid extraction. After the RT-PCR process, the viruses were then captured by the MNPs coated with amino-modified probes and carboxyl. Multiplexed detection was realized with the abilities to detect less than 100 copies of viruses per microliter of serum.

**Table 3. Magnetic Materials as Auxiliary Tools in Other Immunoassay Platforms**

| Platform | Assay Time | Pathogens | Limit of Detection (LOD) | Evaluated Matrices | Role of magnetic materials | Ref. |
| --- | --- | --- | --- | --- | --- | --- |
| MALDI-TOF MS | 1 h | H5N2 | $10^{4.5} - 10^{5.5}$ TCID$_{50}$ | Virus lysate | Magnetic separation | [112] |
| FMR | N.A. | Vibrio cholerae | 1 pM in 1 μL sample | PBS | Magnetic signal source | [113] |
| | | Zika virus | 100 aM in 1 μL sample | PBS containing 20% fetal bovine serum | | |
| SERS | 25 min | H3N2 | $10^2$ TCID$_{50}$/mL in 100 μL sample | PBS | SERS-active magnetic supporting substrates | [114] |
| Fluorescent | 2 h | EIV | 1.3 ng/mL EIV antigen | PBS | Magnetic separation | [116] |
| | | EIAV | 1.2 ng/mL EIAV antigen | | | |
| RLS | 12 h | HAV | 6.2 pM | N.A. | Magnetic separation and magnetic signal source | [117] |
| | 20 min | JEV | 1.3 pM | Serum | Magnetic separation and magnetic signal source | [118] |
| RT-PCR | 2 h | HBV | 10 copies in 1 μL sample | Serum | Magnetic separation | [119] |
| | | HCV | 10 copies in 1 μL sample | | | |
| | | HIV | 100 copies in 1 μL sample | | | |

## 6. Conclusions

To sum it all up, in the midst of COVID-19 pandemic, the demands for high sensitivity, low cost, rapid, easy-to-use, and reliable disease testing tools are increasing. Current diagnostic tests for COVID-19 are based on real time RT-PCR (rRT-PCR) assays. Although it is sensitive, PCR requires expensive equipment, trained technicians to perform the test and have long turnaround times. In addition, its availability is impeded by a shortage in supply during the current emergency. As of July 2020, there is no effective vaccine to prevent the spread of COVID-19. As the globe is searching for effective cures for COVID-19, actions are also being taken to search for better and faster diagnosis tools for timely diagnosis, management, and control the COVID-19. We reviewed the magnetic immunoassay literatures prior to COVID-19 and highlighted some promising tools for detecting pathogens as well as viruses with high specificity and sensitivity. All the detection platforms reviewed in this paper can be extended to other microbial or viral organisms with a change in the specificity of the reagents on MNPs. It is expected that the magnetic immunoassay platforms will transform today's expensive and labor-intensive diagnostic techniques into a user-friendly and cost-effective detection protocol with superior or comparable sensitivity. This paradigm shift could contribute to better surveillance and control of SARS-CoV-2 infection in populations.

In addition, this review paper focuses on magnetic immunoassay platforms for pathogen and virus detections and different detection tools are reported and categorized by different technologies. However, detection platforms can also be categorized by the target biomarkers such as nucleic acid testing and protein testing (protein antigens and antibodies). Other non-magnetic diagnostic tools such as computed tomography (CT) scans and nucleic acid analysis are prevalently used for diagnosing and screening COVID-19.[120–124] In the end, however, a very basic question still lingers in our mind: where are these nanobiosensors when the world is fighting a global health pandemic? Why aren't they being put to commercialization? This can be answered from several point-of-views. From technical view, an ideal biosensor should meet most or all of the following requirements: high sensitivity, high selectivity, quick response time, multiplexing capabilities, multiple sensing modes, disposable, long shelf life and easy-to-use. Pros and cons for the magnetic biosensors given in Table 2 clearly indicate that all technologies lack something or the other from technical point of view. Furthermore, advancements in immunoassay platform require investments from industry for the particular technology to be mass manufacturable, autonomous and cost-effective. So far, magnetic biosensors have not been commercialized to a very large extent. That is why portable magnetic immunoassay platforms haven't been a big shot in the midst of this global pandemic.

**Associated Content**

**ORCID**

Kai Wu: 0000-0002-9444-6112

Renata Saha: 0000-0002-0389-0083

Diqing Su: 0000-0002-5790-8744

Venkatramana D. Krishna: 0000-0002-1980-5525

Jinming Liu: 0000-0002-4313-5816

Maxim C-J Cheeran: 0000-0002-5331-4746

Jian-Ping Wang: 0000-0003-2815-6624

**Notes**

The authors declare no conflict of interest.

**Acknowledgments**

This study was financially supported by the Institute of Engineering in Medicine, the University of Minnesota Medical School and the University of Minnesota Physicians and Fairview Health Services through COVID-19 Rapid Response Grant.

**References**

[1]   Lu H, Stratton C W and Tang Y-W 2020 Outbreak of pneumonia of unknown etiology in Wuhan, China: The mystery and the miracle *J. Med. Virol.* **92** 401–2

[2]  Zhu N, Zhang D, Wang W, Li X, Yang B, Song J, Zhao X, Huang B, Shi W and Lu R 2020 A novel coronavirus from patients with pneumonia in China, 2019 *N. Engl. J. Med.*

[3]  Wu F, Zhao S, Yu B, Chen Y-M, Wang W, Song Z-G, Hu Y, Tao Z-W, Tian J-H and Pei Y-Y 2020 A new coronavirus associated with human respiratory disease in China *Nature* **579** 265–9

[4]  of the International C S G 2020 The species Severe acute respiratory syndrome-related coronavirus: classifying 2019-nCoV and naming it SARS-CoV-2 *Nat. Microbiol.* **5** 536

[5]  Zhou P, Yang X-L, Wang X-G, Hu B, Zhang L, Zhang W, Si H-R, Zhu Y, Li B and Huang C-L 2020 A pneumonia outbreak associated with a new coronavirus of probable bat origin *nature* **579** 270–3

[6]  Chan J F-W, Kok K-H, Zhu Z, Chu H, To K K-W, Yuan S and Yuen K-Y 2020 Genomic characterization of the 2019 novel human-pathogenic coronavirus isolated from a patient with atypical pneumonia after visiting Wuhan *Emerg. Microbes Infect.* **9** 221–36

[7]  Khailany R A, Safdar M and Ozaslan M 2020 Genomic characterization of a novel SARS-CoV-2 *Gene Rep.* 100682

[8]  Sawicki S G, Sawicki D L and Siddell S G 2007 A contemporary view of coronavirus transcription *J. Virol.* **81** 20–9

[9]  Chen N, Zhou M, Dong X, Qu J, Gong F, Han Y, Qiu Y, Wang J, Liu Y and Wei Y 2020 Epidemiological and clinical characteristics of 99 cases of 2019 novel coronavirus pneumonia in Wuhan, China: a descriptive study *The Lancet* **395** 507–13

[10]  Huang C, Wang Y, Li X, Ren L, Zhao J, Hu Y, Zhang L, Fan G, Xu J and Gu X 2020 Clinical features of patients infected with 2019 novel coronavirus in Wuhan, China *The lancet* **395** 497–506

[11]  Wang D, Hu B, Hu C, Zhu F, Liu X, Zhang J, Wang B, Xiang H, Cheng Z and Xiong Y 2020 Clinical characteristics of 138 hospitalized patients with 2019 novel coronavirus–infected pneumonia in Wuhan, China *Jama* **323** 1061–9

[12]  Corman V M, Landt O, Kaiser M, Molenkamp R, Meijer A, Chu D K, Bleicker T, Brünink S, Schneider J and Schmidt M L 2020 Detection of 2019 novel coronavirus (2019-nCoV) by real-time RT-PCR *Eurosurveillance* **25** 2000045

[13]  Chu D K, Pan Y, Cheng S M, Hui K P, Krishnan P, Liu Y, Ng D Y, Wan C K, Yang P and Wang Q 2020 Molecular diagnosis of a novel coronavirus (2019-nCoV) causing an outbreak of pneumonia *Clin. Chem.* **66** 549–55

[14]  Kucirka L M, Lauer S A, Laeyendecker O, Boon D and Lessler J 2020 Variation in false-negative rate of reverse transcriptase polymerase chain reaction–based SARS-CoV-2 tests by time since exposure *Ann. Intern. Med.*

[15]  Qin C, Liu F, Yen T-C and Lan X 2020 18 F-FDG PET/CT findings of COVID-19: a series of four highly suspected cases *Eur. J. Nucl. Med. Mol. Imaging* 1–6

[16] Li D, Wang D, Dong J, Wang N, Huang H, Xu H and Xia C 2020 False-negative results of real-time reverse-transcriptase polymerase chain reaction for severe acute respiratory syndrome coronavirus 2: role of deep-learning-based CT diagnosis and insights from two cases *Korean J. Radiol.* **21** 505–8

[17] Zhao J, Yuan Q, Wang H, Liu W, Liao X, Su Y, Wang X, Yuan J, Li T and Li J 2020 Antibody responses to SARS-CoV-2 in patients of novel coronavirus disease 2019 *Clin. Infect. Dis.*

[18] Arevalo-Rodriguez I, Buitrago-Garcia D, Simancas-Racines D, Zambrano-Achig P, del Campo R, Ciapponi A, Sued O, Martinez-Garcia L, Rutjes A and Low N 2020 False-negative results of initial RT-PCR assays for COVID-19: a systematic review *medRxiv*

[19] Zhen W, Smith E, Manji R, Schron D and Berry G J 2020 Clinical evaluation of three sample-to-Answer platforms for the detection of SARS-CoV-2 *J. Clin. Microbiol.*

[20] Basu A, Zinger T, Inglima K, Woo K, Atie O, Yurasits L, See B and Aguero-Rosenfeld M E 2020 Performance of Abbott ID NOW COVID-19 rapid nucleic acid amplification test in nasopharyngeal swabs transported in viral media and dry nasal swabs, in a New York City academic institution *J. Clin. Microbiol.*

[21] Mitchell S L and George K S 2020 Evaluation of the COVID19 ID NOW EUA Assay *J. Clin. Virol.*

[22] Amanat F, Stadlbauer D, Strohmeier S, Nguyen T H, Chromikova V, McMahon M, Jiang K, Arunkumar G A, Jurczyszak D and Polanco J 2020 A serological assay to detect SARS-CoV-2 seroconversion in humans *Nat. Med.* 1–4

[23] Beavis K G, Matushek S M, Abeleda A P F, Bethel C, Hunt C, Gillen S, Moran A and Tesic V 2020 Evaluation of the EUROIMMUN Anti-SARS-CoV-2 ELISA Assay for detection of IgA and IgG antibodies *J. Clin. Virol.* 104468

[24] Liu R, Liu X, Yuan L, Han H, Shereen M A, Zhen J, Niu Z, Li D, Liu F and Wu K 2020 Analysis of Adjunctive Serological Detection to Nucleic Acid Test for Severe Acute Respiratory Syndrome Coronavirus 2 (SARS-CoV-2) Infection Diagnosis *Int. Immunopharmacol.* 106746

[25] Zhang W, Du R-H, Li B, Zheng X-S, Yang X-L, Hu B, Wang Y-Y, Xiao G-F, Yan B and Shi Z-L 2020 Molecular and serological investigation of 2019-nCoV infected patients: implication of multiple shedding routes *Emerg. Microbes Infect.* **9** 386–9

[26] Lv H, Wu N C, Tsang O T-Y, Yuan M, Perera R A, Leung W S, So R T, Chan J M C, Yip G K and Chik T S H 2020 Cross-reactive antibody response between SARS-CoV-2 and SARS-CoV infections *Cell Rep.* 107725

[27] Su D, Wu K, Krishna V, Klein T, Liu J, Feng Y, Perez A M, Cheeran M C and Wang J-P 2019 Detection of Influenza a Virus in Swine Nasal Swab Samples With a Wash-Free Magnetic Bioassay and a Handheld Giant Magnetoresistance Sensing System *Front. Microbiol.* **10** 1077

[28] Rettcher S, Jungk F, Kühn C, Krause H-J, Nölke G, Commandeur U, Fischer R, Schillberg S and Schröper F 2015 Simple and portable magnetic immunoassay for rapid detection and sensitive quantification of plant viruses *Appl Env. Microbiol* **81** 3039–48

[29] Aytur T, Foley J, Anwar M, Boser B, Harris E and Beatty P R 2006 A novel magnetic bead bioassay platform using a microchip-based sensor for infectious disease diagnosis *J. Immunol. Methods* **314** 21–9

[30] Hash S, Martinez-Viedma M P, Fung F, Han J E, Yang P, Wong C, Doraisamy L, Menon S and Lightner D 2019 Nuclear magnetic resonance biosensor for rapid detection of Vibrio parahaemolyticus *Biomed. J.* **42** 187–92

[31] Lin H, Lu Q, Ge S, Cai Q and Grimes C A 2010 Detection of pathogen Escherichia coli O157: H7 with a wireless magnetoelastic-sensing device amplified by using chitosan-modified magnetic Fe3O4 nanoparticles *Sens. Actuators B Chem.* **147** 343–9

[32] Zou D, Jin L, Wu B, Hu L, Chen X, Huang G and Zhang J 2019 Rapid detection of Salmonella in milk by biofunctionalised magnetic nanoparticle cluster sensor based on nuclear magnetic resonance *Int. Dairy J.* **91** 82–8

[33] Gao Y, Huo W, Zhang L, Lian J, Tao W, Song C, Tang J, Shi S and Gao Y 2019 Multiplex measurement of twelve tumor markers using a GMR multi-biomarker immunoassay biosensor *Biosens. Bioelectron.* **123** 204–10

[34] Wang T, Yang Z, Lei C, Lei J and Zhou Y 2014 An integrated giant magnetoimpedance biosensor for detection of biomarker *Biosens. Bioelectron.* **58** 338–44

[35] Klein T, Wang W, Yu L, Wu K, Boylan K L, Vogel R I, Skubitz A P and Wang J-P 2019 Development of a multiplexed giant magnetoresistive biosensor array prototype to quantify ovarian cancer biomarkers *Biosens. Bioelectron.* **126** 301–7

[36] Wang W, Wang Y, Tu L, Klein T, Feng Y, Li Q and Wang J-P 2014 Magnetic detection of mercuric ion using giant magnetoresistance-based biosensing system *Anal. Chem.* **86** 3712–6

[37] Schotter J, Kamp P B, Becker A, Puhler A, Reiss G and Bruckl H 2004 Comparison of a prototype magnetoresistive biosensor to standard fluorescent DNA detection *Biosens. Bioelectron.* **19** 1149–56

[38] Thomson W 1857 XIX. On the electro-dynamic qualities of metals:—Effects of magnetization on the electric conductivity of nickel and of iron *Proc. R. Soc. Lond.* 546–50

[39] McGuire T and Potter R L 1975 Anisotropic magnetoresistance in ferromagnetic 3d alloys *IEEE Trans. Magn.* **11** 1018–38

[40] Fert A and Campbell I A 1976 Electrical resistivity of ferromagnetic nickel and iron based alloys *J. Phys. F Met. Phys.* **6** 849

[41] Baibich M N, Broto J M, Fert A, Van Dau F N, Petroff F, Etienne P, Creuzet G, Friederich A and Chazelas J 11 Giant Magnetoresistance of (001)Fe/(001)Cr Magnetic Superlattices *Phys. Rev. Lett.* **61** 2472–5

[42] Binasch G, Grünberg P, Saurenbach F and Zinn W January 3 Enhanced magnetoresistance in layered magnetic structures with antiferromagnetic interlayer exchange *Phys. Rev. B* **39** 4828–30

[43] Parkin S, Jiang X, Kaiser C, Panchula A, Roche K and Samant M 2003 Magnetically engineered spintronic sensors and memory *Proc. IEEE* **91** 661–80

[44] Bernardi J, Hütten A and Thomas G 1996 Electron microscopy of giant magnetoresistive granular Au–Co alloys *J. Magn. Magn. Mater.* **157** 153–5

[45] Hütten A, Bernardi J, Friedrichs S, Thomas G and Balcells L 1995 Microstructural influence on magnetic properties and giant magnetoresistance of melt-spun gold-cobalt *Scr. Metall. Mater.* **33** 1647–66

[46] Garcia-Torres J, Vallés E and Gómez E 2011 Giant magnetoresistance in electrodeposited Co–Ag granular films *Mater. Lett.* **65** 1865–7

[47] Kenane S, Voiron J, Benbrahim N, Chaînet E and Robaut F 2006 Magnetic properties and giant magnetoresistance in electrodeposited Co–Ag granular films *J. Magn. Magn. Mater.* **297** 99–106

[48] Odnodvorets L V, Protsenko I Y, Tkach O P, Shabelnyk Y M and Shumakova N I 2017 Magnetoresistive Effect in Granular Film Alloys Based on Ag and Fe or Co *J. Nano- Electron. Phys.* **9** 2021–1

[49] Melzer M, Karnaushenko D, Makarov D, Baraban L, Calvimontes A, Mönch I, Kaltofen R, Mei Y and Schmidt O G 2012 Elastic magnetic sensor with isotropic sensitivity for in-flow detection of magnetic objects *Rsc Adv.* **2** 2284–8

[50] Melzer M, Karnaushenko D, Lin G, Baunack S, Makarov D and Schmidt O G 2015 Direct transfer of magnetic sensor devices to elastomeric supports for stretchable electronics *Adv. Mater.* **27** 1333–8

[51] Noh E-S, Ulloa S E and Lee H-M 2007 A theoretical study of an amorphous aluminium oxide layer used as a tunnel barrier in a magnetic tunnel junction *Phys. Status Solidi B* **244** 4427–30

[52] Gloos K, Koppinen P J and Pekola J P 2003 Properties of native ultrathin aluminium oxide tunnel barriers *J. Phys. Condens. Matter* **15** 1733

[53] Parkin S S P, Kaiser C, Panchula A, Rice P M, Hughes B, Samant M and Yang S H 2004 Giant tunnelling magnetoresistance at room temperature with MgO (100) tunnel barriers *Nat. Mater.* **3** 862–7

[54] Yuasa S, Fukushima A, Nagahama T, Ando K and Suzuki Y 2004 High tunnel magnetoresistance at room temperature in fully epitaxial Fe/MgO/Fe tunnel junctions due to coherent spin-polarized tunneling *Jpn. J. Appl. Phys.* **43** L588

[55] Ferreira R, Paz E, Freitas P P, Wang J and Xue S 2012 Large area and low aspect ratio linear magnetic tunnel junctions with a soft-pinned sensing layer *IEEE Trans. Magn.* **48** 3719–22

[56] Chen J Y, Feng J F and Coey J M D 2012 Tunable linear magnetoresistance in MgO magnetic tunnel junction sensors using two pinned CoFeB electrodes *Appl. Phys. Lett.* **100** 142407

[57] Cao J and Freitas P P 2010 Wheatstone bridge sensor composed of linear MgO magnetic tunnel junctions *J. Appl. Phys.* **107** 09E712

[58] Nowak E R, Weissman M B and Parkin S S P 1999 Electrical noise in hysteretic ferromagnet–insulator–ferromagnet tunnel junctions *Appl. Phys. Lett.* **74** 600–2

[59] Wu K, Su D, Liu J, Saha R and Wang J-P 2019 Magnetic nanoparticles in nanomedicine: a review of recent advances *Nanotechnology* **30** 502003

[60] Moretti D, DiFrancesco M L, Sharma P P, Dante S, Albisetti E, Monticelli M, Bertacco R, Petti D, Baldelli P and Benfenati F 2018 Biocompatibility of a Magnetic Tunnel Junction sensor array for the detection of neuronal signals in culture *Front. Neurosci.* **12** 909

[61] Su D, Wu K, Saha R, Peng C and Wang J-P 2020 Advances in Magnetoresistive Biosensors *Micromachines* **11** 34

[62] Baselt D R, Lee G U, Natesan M, Metzger S W, Sheehan P E and Colton R J 1998 A biosensor based on magnetoresistance technology *Biosens. Bioelectron.* **13** 731–9

[63] Krishna V D, Wu K, Perez A M and Wang J P 2016 Giant Magnetoresistance-based Biosensor for Detection of Influenza A Virus *Front. Microbiol.* **7** 8

[64] Wu K, Klein T, Krishna V D, Su D, Perez A M and Wang J-P 2017 Portable GMR Handheld Platform for the Detection of Influenza A Virus *ACS Sens.* **2** 1594–601

[65] Choi J, Gani A W, Bechstein D J, Lee J-R, Utz P J and Wang S X 2016 Portable, one-step, and rapid GMR biosensor platform with smartphone interface *Biosens. Bioelectron.* **85** 1–7

[66] Zhi X, Liu Q S, Zhang X, Zhang Y X, Feng J and Cui D X 2012 Quick genotyping detection of HBV by giant magnetoresistive biochip combined with PCR and line probe assay *Lab. Chip* **12** 741–5

[67] Zhi X, Deng M, Yang H, Gao G, Wang K, Fu H, Zhang Y, Chen D and Cui D 2014 A novel HBV genotypes detecting system combined with microfluidic chip, loop-mediated isothermal amplification and GMR sensors *Biosens. Bioelectron.* **54** 372–7

[68] Sun X, Lei C, Guo L and Zhou Y 2016 Separable detecting of Escherichia coli O157H: H7 by a giant magneto-resistance-based bio-sensing system *Sens. Actuators B Chem.* **234** 485–92

[69] Gupta S and Kakkar V 2019 DARPin based GMR Biosensor for the detection of ESAT-6 Tuberculosis Protein *Tuberculosis* **118** 101852

[70] Grancharov S G, Zeng H, Sun S, Wang S X, O'Brien S, Murray C, Kirtley J and Held G 2005 Bio-functionalization of monodisperse magnetic nanoparticles and their use as biomolecular labels in a magnetic tunnel junction based sensor *J. Phys. Chem. B* **109** 13030–5

[71] Shen W, Schrag B D, Carter M J and Xiao G 2008 Quantitative detection of DNA labeled with magnetic nanoparticles using arrays of MgO-based magnetic tunnel junction sensors *Appl. Phys. Lett.* **93** 033903

[72] Shen W, Schrag B D, Carter M J, Xie J, Xu C, Sun S and Xiao G 2008 Detection of DNA labeled with magnetic nanoparticles using MgO-based magnetic tunnel junction sensors *J. Appl. Phys.* **103** 7A306

[73] Gervasoni G, Carminati M, Ferrari G, Sampietro M, Albisetti E, Petti D, Sharma P and Bertacco R 2014 A 12-channel dual-lock-in platform for magneto-resistive DNA detection with ppm resolution *2014 IEEE Biomedical Circuits and Systems Conference (BioCAS) Proceedings* (IEEE) pp 316–9

[74] Sharma P P, Albisetti E, Massetti M, Scolari M, La Torre C, Monticelli M, Leone M, Damin F, Gervasoni G, Ferrari G, Salice F, Cerquaglia E, Falduti G, Cretich M, Marchisio E, Chiari M, Sampietro M, Petti D and Bertacco R 2017 Integrated platform for detecting pathogenic DNA via magnetic tunneling junction-based biosensors *Sens. Actuators B-Chem.* **242** 280–7

[75] Li L, Mak K Y and Zhou Y 2020 Detection of HIV-1 antigen based on magnetic tunnel junction sensor and magnetic nanoparticles *Chin. Phys. B*

[76] Orlov A V, Znoyko S L, Cherkasov V R, Nikitin M P and Nikitin P I 2016 Multiplex biosensing based on highly sensitive magnetic nanolabel quantification: rapid detection of botulinum neurotoxins A, B, and E in liquids *Anal. Chem.* **88** 10419–26

[77] Orlov A V, Khodakova J A, Nikitin M P, Shepelyakovskaya A O, Brovko F A, Laman A G, Grishin E V and Nikitin P I 2013 Magnetic Immunoassay for Detection of Staphylococcal Toxins in Complex Media *Anal. Chem.* **85** 1154–63

[78] Wu K, Liu J, Saha R, Su D, Krishna V D, Cheeran M C-J and Wang J-P 2020 Magnetic Particle Spectroscopy for Detection of Influenza A Virus Subtype H1N1 *ACS Appl. Mater. Interfaces* **12** 13686–97

[79] Pietschmann J, Voepel N, Spiegel H, Krause H-J and Schroeper F 2020 Brief Communication: Magnetic Immuno-Detection of SARS-CoV-2 specific Antibodies *bioRxiv*

[80] Alocilja E C and Luo Y 2017 Portable nuclear magnetic resonance biosensor and assay for a highly sensitive and rapid detection of foodborne bacteria in complex matrices *J. Biol. Eng.* **11** 14

[81] Liong M, Hoang A N, Chung J, Gural N, Ford C B, Min C, Shah R R, Ahmad R, Fernandez-Suarez M and Fortune S M 2013 Magnetic barcode assay for genetic detection of pathogens *Nat. Commun.* **4** 1–9

[82] Nikitin P I, Vetoshko P M and Ksenevich T I 2007 New type of biosensor based on magnetic nanoparticle detection *J. Magn. Magn. Mater.* **311** 445–9

[83] Krause H-J, Wolters N, Zhang Y, Offenhäusser A, Miethe P, Meyer M H, Hartmann M and Keusgen M 2007 Magnetic particle detection by frequency mixing for immunoassay applications *J. Magn. Magn. Mater.* **311** 436–44

[84] Gleich B and Weizenecker J 2005 Tomographic imaging using the nonlinear response of magnetic particles *Nature* **435** 1214–7

[85] Wu K, Su D, Saha R, Wong D and Wang J-P 2019 Magnetic particle spectroscopy-based bioassays: methods, applications, advances, and future opportunities *J. Phys. Appl. Phys.* **52** 173001

[86] Zhang X, Reeves D B, Perreard I M, Kett W C, Griswold K E, Gimi B and Weaver J B 2013 Molecular sensing with magnetic nanoparticles using magnetic spectroscopy of nanoparticle Brownian motion *Biosens. Bioelectron.* **50** 441–6

[87] Khurshid H, Shi Y, Berwin B L and Weaver J B 2018 Evaluating blood clot progression using magnetic particle spectroscopy *Med. Phys.* **45** 3258–63

[88] Znoyko S L, Orlov A V, Bragina V A, Nikitin M P and Nikitin P I 2020 Nanomagnetic lateral flow assay for high-precision quantification of diagnostically relevant concentrations of serum TSH *Talanta* 120961

[89] Znoyko S L, Orlov A V, Pushkarev A V, Mochalova E N, Guteneva N V, Lunin A V, Nikitin M P and Nikitin P I 2018 Ultrasensitive quantitative detection of small molecules with rapid lateral-flow assay based on high-affinity bifunctional ligand and magnetic nanolabels *Anal. Chim. Acta* **1034** 161–7

[90] Wu K, Liu J, Su D, Saha R and Wang J-P 2019 Magnetic Nanoparticle Relaxation Dynamics-Based Magnetic Particle Spectroscopy for Rapid and Wash-Free Molecular Sensing *ACS Appl. Mater. Interfaces* **11** 22979–86

[91] Shi Y, Jyoti D, Gordon-Wylie S W and Weaver J B 2020 Quantification of magnetic nanoparticles by compensating for multiple environment changes simultaneously *Nanoscale* **12** 195–200

[92] Bragina V A, Znoyko S L, Orlov A V, Pushkarev A V, Nikitin M P and Nikitin P I 2019 Analytical platform with selectable assay parameters based on three functions of magnetic nanoparticles: demonstration of highly sensitive rapid quantitation of staphylococcal enterotoxin B in food *Anal. Chem.* **91** 9852–7

[93] Guteneva N V, Znoyko S L, Orlov A V, Nikitin M P and Nikitin P I 2019 Rapid lateral flow assays based on the quantification of magnetic nanoparticle labels for multiplexed immunodetection of small molecules: application to the determination of drugs of abuse *Microchim. Acta* **186** 621

[94] Wu K, Tu L, Su D and Wang J-P 2017 Magnetic dynamics of ferrofluids: mathematical models and experimental investigations *J. Phys. Appl. Phys.* **50** 085005

[95] Wu K, Su D, Saha R, Liu J and Wang J-P 2019 Investigating the effect of magnetic dipole–dipole interaction on magnetic particle spectroscopy: implications for magnetic nanoparticle-based bioassays and magnetic particle imaging *J. Phys. Appl. Phys.* **52** 335002

[96] Draack S, Viereck T, Kuhlmann C, Schilling M and Ludwig F 2017 Temperature-dependent MPS measurements *Int. J. Magn. Part. Imaging* **3**

[97] Perreard I, Reeves D, Zhang X, Kuehlert E, Forauer E and Weaver J 2014 Temperature of the magnetic nanoparticle microenvironment: estimation from relaxation times *Phys. Med. Biol.* **59** 1109

[98] Wu K, Liu J, Wang Y, Ye C, Feng Y and Wang J-P 2015 Superparamagnetic nanoparticle-based viscosity test *Appl. Phys. Lett.* **107** 053701

[99] Utkur M, Muslu Y and Saritas E U 2017 Relaxation-based viscosity mapping for magnetic particle imaging *Phys. Med. Biol.* **62** 3422

[100]   Liu J, Su D, Wu K and Wang J-P 2020 High-moment magnetic nanoparticles *J. Nanoparticle Res.* **22** 1–16

[101]   Anfossi L, Di Nardo F, Russo A, Cavalera S, Giovannoli C, Spano G, Baumgartner S, Lauter K and Baggiani C 2019 Silver and gold nanoparticles as multi-chromatic lateral flow assay probes for the detection of food allergens *Anal. Bioanal. Chem.* **411** 1905–13

[102]   Ruppert C, Phogat N, Laufer S, Kohl M and Deigner H-P 2019 A smartphone readout system for gold nanoparticle-based lateral flow assays: application to monitoring of digoxigenin *Microchim. Acta* **186** 119

[103]   Wang C, Xiao R, Wang S, Yang X, Bai Z, Li X, Rong Z, Shen B and Wang S 2019 Magnetic quantum dot based lateral flow assay biosensor for multiplex and sensitive detection of protein toxins in food samples *Biosens. Bioelectron.* **146** 111754

[104]   Wang J, Meng H-M, Chen J, Liu J, Zhang L, Qu L, Li Z and Lin Y 2019 Quantum dot-based lateral flow test strips for highly sensitive detection of the tetanus antibody *ACS Omega* **4** 6789–95

[105]   Zhu M, Jia Y, Peng L, Ma J, Li X and Shi F 2019 A highly sensitive dual-color lateral flow immunoassay for brucellosis using one-step synthesized latex microspheres *Anal. Methods* **11** 2937–42

[106]   Issadore D, Min C, Liong M, Chung J, Weissleder R and Lee H 2011 Miniature magnetic resonance system for point-of-care diagnostics *Lab. Chip* **11** 2282–7

[107]   Viegas A, Manso J, Nobrega F L and Cabrita E J 2011 Saturation-transfer difference (STD) NMR: a simple and fast method for ligand screening and characterization of protein binding *J. Chem. Educ.* **88** 990–4

[108]   Yue D, Zhang Y, Cheng L, Ma J, Xi Y, Yang L, Su C, Shao B, Huang A and Xiang R 2016 Hepatitis B virus X protein (HBx)-induced abnormalities of nucleic acid metabolism revealed by 1 H-NMR-based metabonomics *Sci. Rep.* **6** 1–13

[109]   Kusunoki H, Tanaka T, Kohno T, Kimura H, Hosoda K, Wakamatsu K and Hamaguchi I 2019 NMR characterization of the interaction between Bcl-xL and the BH3-like motif of hepatitis B virus X protein *Biochem. Biophys. Res. Commun.* **518** 445–50

[110]   Dapiaggi F, Pieraccini S, Potenza D, Vasile F, Macut H, Pellegrino S, Aliverti A and Sironi M 2017 Computer aided design and NMR characterization of an oligopeptide targeting the Ebola virus VP24 protein *New J. Chem.* **41** 4308–15

[111]   Vasile F, Gubinelli F, Panigada M, Soprana E, Siccardi A and Potenza D 2018 NMR interaction studies of Neu5Ac-α-(2, 6)-Gal-β-(1-4)-GlcNAc with influenza-virus hemagglutinin expressed in transfected human cells *Glycobiology* **28** 42–9

[112]   Chou T-C, Hsu W, Wang C-H, Chen Y-J and Fang J-M 2011 Rapid and specific influenza virus detection by functionalized magnetic nanoparticles and mass spectrometry *J. Nanobiotechnology* **9** 52

[113] Tian B, Liao X, Svedlindh P, Strömberg M and Wetterskog E 2018 Ferromagnetic resonance biosensor for homogeneous and volumetric detection of DNA *ACS Sens.* **3** 1093–101

[114] Sun Y, Xu L, Zhang F, Song Z, Hu Y, Ji Y, Shen J, Li B, Lu H and Yang H 2017 A promising magnetic SERS immunosensor for sensitive detection of avian influenza virus *Biosens. Bioelectron.* **89** 906–12

[115] Barrios-Gumiel A, Sepúlveda-Crespo D, Jiménez J L, Gómez R, Muñoz-Fernández M Á and de la Mata F J 2019 Dendronized magnetic nanoparticles for HIV-1 capture and rapid diagnostic *Colloids Surf. B Biointerfaces* **181** 360–8

[116] Wang G, Gao Y, Huang H and Su X 2010 Multiplex immunoassays of equine virus based on fluorescent encoded magnetic composite nanoparticles *Anal. Bioanal. Chem.* **398** 805–13

[117] Zhang F, Luo L, Gong H, Chen C and Cai C 2018 A magnetic molecularly imprinted optical chemical sensor for specific recognition of trace quantities of virus *RSC Adv.* **8** 32262–8

[118] Luo L, Yang J, Liang K, Chen C, Chen X and Cai C 2019 Fast and sensitive detection of Japanese encephalitis virus based on a magnetic molecular imprinted polymer–resonance light scattering sensor *Talanta* **202** 21–6

[119] Ali Z, Wang J, Tang Y, Liu B, He N and Li Z 2017 Simultaneous detection of multiple viruses based on chemiluminescence and magnetic separation *Biomater. Sci.* **5** 57–66

[120] Li Y and Xia L 2020 Coronavirus disease 2019 (COVID-19): role of chest CT in diagnosis and management *Am. J. Roentgenol.* **214** 1280–6

[121] Guo L, Ren L, Yang S, Xiao M, Chang D, Yang F, Dela Cruz C S, Wang Y, Wu C and Xiao Y 2020 Profiling early humoral response to diagnose novel coronavirus disease (COVID-19) *Clin. Infect. Dis.*

[122] Chan J F-W, Yip C C-Y, To K K-W, Tang T H-C, Wong S C-Y, Leung K-H, Fung A Y-F, Ng A C-K, Zou Z and Tsoi H-W 2020 Improved molecular diagnosis of COVID-19 by the novel, highly sensitive and specific COVID-19-RdRp/Hel real-time reverse transcription-PCR assay validated in vitro and with clinical specimens *J. Clin. Microbiol.* **58**

[123] Zhai P, Ding Y, Wu X, Long J, Zhong Y and Li Y 2020 The epidemiology, diagnosis and treatment of COVID-19 *Int. J. Antimicrob. Agents* 105955

[124] Long C, Xu H, Shen Q, Zhang X, Fan B, Wang C, Zeng B, Li Z, Li X and Li H 2020 Diagnosis of the Coronavirus disease (COVID-19): rRT-PCR or CT? *Eur. J. Radiol.* 108961